\documentclass{aa}  
\usepackage{graphicx}	
\usepackage{amsmath}	
\usepackage{amssymb}	
\usepackage{bm}	        
\usepackage{csvsimple}
\usepackage{ragged2e}
\usepackage{multirow}
\usepackage{siunitx} 
\usepackage[utf8]{inputenc}
\usepackage{dirtytalk}
\usepackage{xparse}
\usepackage{enumitem} 
\usepackage{array}
\usepackage{CJKutf8}
\usepackage[flushleft]{threeparttable}
\usepackage{txfonts}
\usepackage[colorlinks, citecolor=blue, linkcolor=blue]{hyperref}
\usepackage{graphicx}
\usepackage{txfonts}
\usepackage{svg}
\usepackage{orcidlink}

\newcommand{\orcid}[1]{\orcidlink{#1}}

\begin{document}

   \title{Emergence of high-mass stars in complex fiber networks (EMERGE)}


   \subtitle{V. From filaments to spheroids: the origin of the hub-filament systems}

\titlerunning{EMERGE (V). From filaments to spheroids: the origin of the hub-filament systems}

   \author{A. Hacar
          \inst{1} \orcid{0000-0001-5397-6961}
          \and
          R. Konietzka \inst{2,3}\orcid{0000-0001-8235-2939}
          \and
          D. Seifried \inst{4}
          \and
          S. E. Clark\inst{5,6}\orcid{0000-0002-7633-3376}
          \and
          A. Socci \inst{1}
          \and 
          F. Bonanomi \inst{1}
          \and
          A. Burkert \inst{7,8,9}
          \and 
          E. Schisano \inst{10}\orcid{0000-0003-1560-3958}
          \and
          J. Kainulainen \inst{11}
          \and
          R. Smith \inst{12}
    }

   \institute{Department of Astrophysics, University of Vienna,
              T\"urkenschanzstrasse 17, A-1180 Vienna\\
              \email{alvaro.hacar@univie.ac.at}
         \and
         Center for Astrophysics $\mid$ Harvard \& Smithsonian, 60 Garden St., Cambridge, MA 02138, USA
         \and
         Department of Astronomy, Harvard University, 60 Garden St., Cambridge, MA 02138, USA
         \and 
         University of Cologne, I. Physical Institute, Z\"ulpicher Str. 77, 50937 Cologne, Germany
         \and 
         Department of Physics, Stanford University, Stanford, CA 94305, USA
        \and
        Kavli Institute for Particle Astrophysics \& Cosmology, P.O. Box 2450, Stanford University, Stanford, CA 94305, USA
        \and
        Universit\"ats-Sternwarte, Ludwig-Maximilians-Universit\"at M\"unchen, Scheinerstrasse 1, D-81679 Munich, Germany 
        \and
        Excellence Cluster ORIGINS, Boltzmannstrasse 2, D-85748 Garching, Germany
        \and
        Max-Planck Institute for Extraterrestrial Physics, Giessenbachstrasse 1, D-85748 Garching, Germany
        \and 
        Istituto di Astrofisica e Planetologia Spaziali (INAF-IAPS), Via Fosso del Cavaliere 100, I-00133, Rome, Italy
        \and
        Chalmers University of Technology, Department of Space, Earth and Environment, SE-412 93, Gothenburg, Sweden
        \and 
        SUPA, School of Physics and Astronomy, University of St Andrews, North Haugh, St Andrews, KY16 9SS, UK
             }

   \date{Received XXX ; accepted XXX}

 
  \abstract
   {Identified as parsec-size, gas clumps at the junction of multiple filaments, hub-filament systems (HFS) play a crucial role during the formation of young clusters and high-mass stars. These HFS appear nevertheless to be detached from most galactic filaments when compared in the mass-length (M-L) phase-space.}
   {We aim to characterize the early evolution of HFS as part of the filamentary description of the interstellar medium.}
   {Combining previous scaling relations with new analytic calculations, we created a toy model to explore the different physical regimes described by the M-L diagram.
   Despite its simplicity, our model accurately reproduces several observational properties reported for filaments and HFS such as their expected typical aspect ratio ($A$), mean surface density ($\Sigma$), and gas accretion rate ($\dot{m}$). Moreover, this model naturally explains the different mass and length regimes populated by filaments and HFS, respectively.
   }
   {Our model predicts a dichotomy between filamentary ($A\geq 3$) and spheroidal ($A<3$) structures connected to the relative importance of their fragmentation, accretion, and collapse timescales.
   Individual filaments with low accretion rates are dominated by an efficient internal fragmentation. In contrast,
   the formation of compact HFS at the intersection of filaments triggers a geometric phase-transition leading to the gravitational collapse of these structures at parsec-scales in $\sim$~1~Myr  also inducing higher accretion rates.
   }
   {}


   \maketitle
%

\section{Introduction}
The filamentary nature of the interstellar medium (ISM) has been recognized since more than a century ago \citep{Barnard1907,Schneider1979}.
Filaments at different scales are formed as part of the hierarchical gas organization of the ISM \citep[see][for a discussion]{Hacar2022}. Organized in complex networks, filaments also play a key role in the ISM evolution promoting the conditions for star formation inside molecular clouds \citep[see][for recent reviews]{Andre2014,Pineda2023}. Characterizing the origin and evolution of this filamentary organization of the ISM is therefore of paramount importance for our current description of the star formation process in our Galaxy.

The so-called hub-filament systems (HFS) define the central massive clumps found at the apparent junction (nodes) of multiple filamentary structures radiating away from its centre. 
HFS were first identified in both nearby and galactic plane star-forming regions \citep{Myers2009a}. After this seminal work, HFS have been systematically found in high-mass star-forming regions and infrared dark clouds \citep[e.g.,][]{Schneider2010,GalvanMadrid2010,Hennemann2012,Peretto2014,Dewangan2020,Liu2021,Panja2023}. HFS are characterized in multiple galactic \citep{Kumar2020,Peretto2022,Morii2023} and extragalactic \citep{Tokuda2023} surveys, as well as dedicated, high-resolution studies \citep{Peretto2013,Williams2018,Hacar2017a,Lu2018,Trevino-Morales2019,Anderson2021,Zhou2024}. Classical HFS show characteristic full width at half maximum $FWHM$ (=$2\times R$, with $R$ as radius) of $\sim$~1~pc, total masses of $>100$~M$_\odot$, and peak gas column densities of $\Sigma > 10^{22}$~cm$^{-2}$ \citep[see][and references therein]{Myers2009a}. Analogous arrangements of filaments are also found at sub-parsec scales  \citep{Wiseman1998,Hacar2018,Zhang2020,Cao2021,Chung2021,Liu2023,AlvarezGutierrez2024} suggesting a self-similar organization of the gas at different scales.

HFS are prime locations for star formation. Young clusters and high-mass stars are preferentially found at the centre of these systems according to recent studies \citep{Myers2009a,Schneider2012,Peretto2013,Tige2017,Motte2018}. Likewise, the most massive clumps in the Galactic Plane appear to be located at the center of HFS \citep{Kumar2020}.
In all cases, HFS show higher column densities than their surrounding filaments.
Gas inflows along multiple filaments feeding their central HFS lead to total mass accretion rates of $\dot{m}\sim 50-2500$~M$_\odot$~pc$^{-1}$~Myr$^{-1}$ \citep{Kirk2013,Peretto2013,Hacar2017a,Hu2021,Zhou2022}, much higher than those observed in individual filaments at similar scales \citep[with $\dot{m}\sim 10-100$~M$_\odot$~pc$^{-1}$~Myr$^{-1}$; see][]{Palmeirim2013,Bonne2020,Schisano2020}. These enhanced accretion rates appear crucial for the fast formation of massive clusters, harbouring high-mass stars, on timescales of $\sim$~1~Myr.

From an observational perspective, HFS are characterized by a low aspect ratio ($A$) \citep{Myers2009a}, usually estimated from the ratio of the longest and shortest FWHM obtained from a 
2D-Gaussian fit to the dust column density distribution \citep[e.g.,][]{Kumar2020}. Similarly, the aspect ratio in filaments is determined by the relation between their length $L$ and radial $FWHM$, i.e., $A=L/FWHM$ \citep[e.g.,][]{Arzoumanian2019}. While filaments usually show large $A\geq 3$ \citep{Arzoumanian2019,Schisano2020}, HFS exhibit more prolate shapes, typically characterized by aspect ratios of $A= 1.2\pm0.4$ \citep{Kumar2020}.
Most of the HFS identification at parsec scales has been carried out either visually on optical/IR images \citep{Myers2009a,Anderson2021} or extracted from the location of nodes identified by filament finding algorithms \citep{Kumar2020}. Novel attempts to characterize these HFS include a quantification of the number of converging filaments per object in a more systematic way \citep{Peretto2022}. 

Despite their key role in star formation, the origin of these HFS is yet unclear. Multiple formation mechanisms were proposed over the years including the regular fragmentation of a compressed gas layer \citep{Myers1983}, { converging flows \citep{Schneider2010,GalvanMadrid2010}, cloud-cloud collisions  \citep{Duarte2010,Duarte2011,Nakamura2014} and/or between filaments at different scales  \citep{Nakamura2012,Kirk2013,Peretto2014,Clarke2017,Hacar2018,Kumar2020,Hoemann2021,Hoemann2024}, and as result of the global hierarchical collapse of molecular clouds \citep{VazquezSemadeni2017,VazquezSemadenietal2019,Camacho2023}. Regardless of their origin, the central clumps in these HFS appear to evolve differently than their radiating filaments. Molecular line observations reveal signatures of gravitational collapse on scales of a parsec as denoted by the presence of converging linear \citep{Kirk2013,Peretto2013,AlvarezGutierrez2024,Sen2024} and accelerated \citep{Hacar2017a,Williams2018,Zhou2022,Zhou2023,Zhang2024_HFS} velocity gradients in the gas flowing along the filaments feeding the central HFS. The transition between filaments and HFS seems however smooth in column density \citep{Peretto2022,Zhou2022}.

In this paper we characterize the physical processes leading to the formation and early evolution of HFS as part of the filamentary structure of the ISM.
We discuss how recent studies have pointed out the apparent dichotomy between the observed properties of isolated filaments (Sect.~\ref{sec:scaling}) and HFS (Sect.~\ref{sec:HFSclusters}). Using a simplified toy model (Sect.~\ref{sec:AR_col}) we demonstrate that the observed properties of HFS (mass, length, aspect ratio, accretion rates, and evolutionary timescales) may be explained by the change from a filamentary to a spheroidal geometry during the ISM evolution at parsec-scales (Sects.~\ref{sec:observations}-\ref{sec:timescales}). This phase-transition induces the gravitational collapse of these HFS and triggers the large accretion rates observed in these objects (Sects.~\ref{sec:inducingcollapse} and \ref{sec:evolution}).

This work is part of the {\it Emergence of high-mass stars in complex fiber networks}\footnote{EMERGE Project website: \url{https://emerge.univie.ac.at/}} (EMERGE) series \citep[Paper I; ][]{Hacar2024}. 
Previous papers (II-IV) of this series describe the internal gas substructure in HFS such as the OMC-1 down to 2000~au resolution using ALMA observations \citep{BonanomiPaperII,SocciPaperIII,SocciPaperIV}. Of particularly interest for our project, in this new work (paper V) we aim to explore the physical origin of HFS at parsec scales and provide direct theoretical predictions for future observations.

\section{From filaments to spheroids}\label{sec:toymodel}

\begin{figure*}[t]
\centering
\includegraphics[width=1.\textwidth]{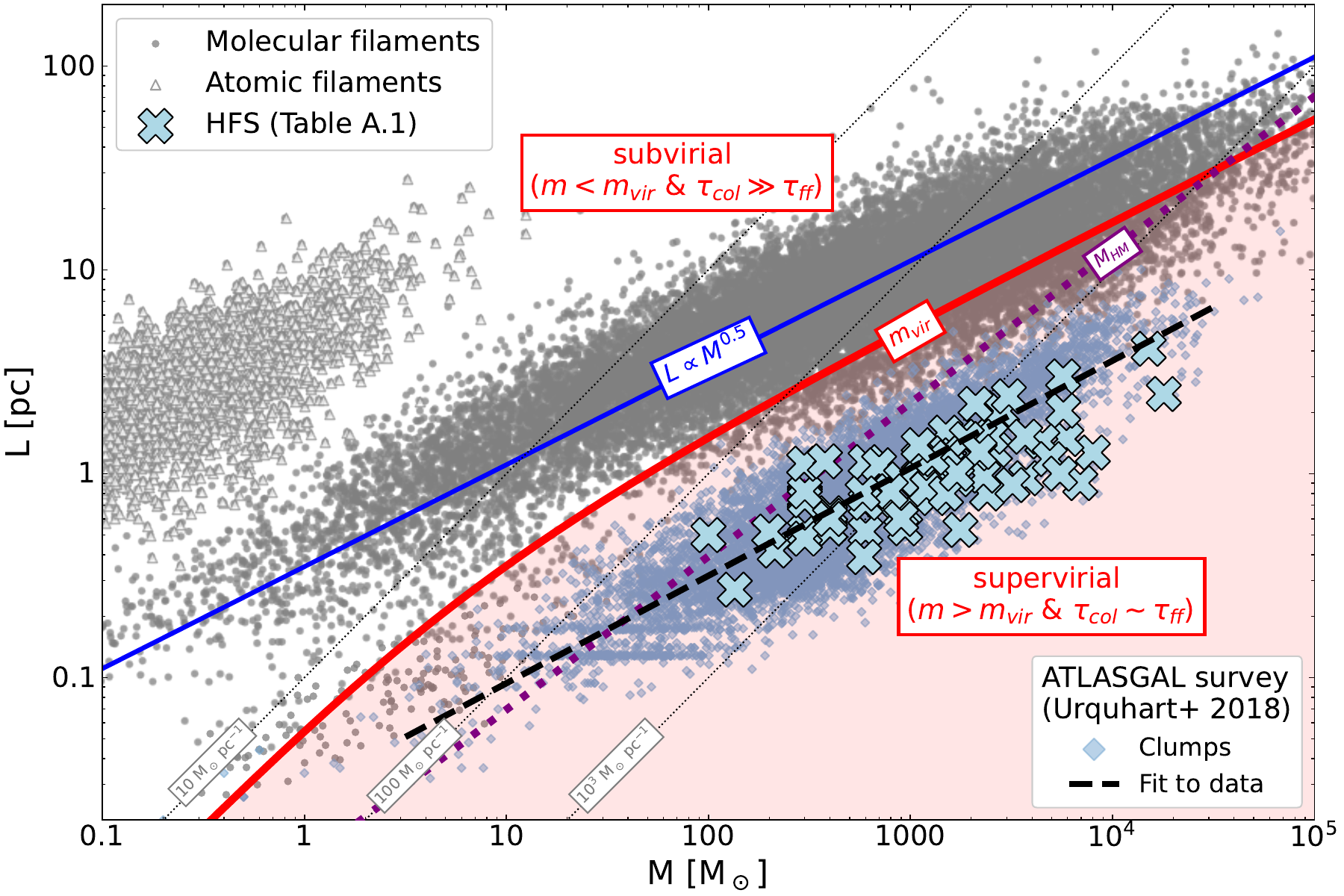}
      \caption{Distribution of molecular (grey dots) and atomic (HI) filaments (grey triangles) in the M-L phase-space. Most data are taken from all the public surveys included in \citet{Hacar2022} (and references therein) plus some additional values from \citet{SocciPaperIII}. Superposed to our data, we also display the M-L distribution of ATLASGAL clumps \citep{Urquhart2018} (light blue diamonds; with $L=2\times R_{AGAL}$ in our notation) and their corresponding power-law fit (dashed black line).
      HFS are also highlighted in this plot (blue crosses; Table~\ref{table:HFS_data}).
      Overall, most filaments follow a M-L scaling relation such as $L\propto M^{0.5}$ (dotted blue line).
      We indicate the virial line mass $m_{vir}$ for equipartition reported for filaments which separates the subvirial and supervirial regimes (solid red line; Eq.~\ref{eq:mvir}) as well as lines of constant line mass $m=[10,100,10^3]$~M$_\odot$~pc$^{-1}$ (dotted grey lines).
      We also include the mass threshold for high-mass star formation $M_{HM}=870\ M_\odot\cdot (R_{eff}/pc)^{1.33}$ \citep[dotted purple line;][]{KauffmannPillai2010}, for which we assumed $L=2\times R_{eff}$.
      }
\label{fig:ML_phasespace}
\end{figure*}

\subsection{Filamentary ISM: scaling relations}\label{sec:scaling}

In a recent meta-analysis, \citet{Hacar2022} have reviewed the physical properties of the filamentary ISM using a comprehensive catalog of more than 22\,000 Galactic filaments collected from 49 independent works using different observational techniques and extraction algorithms. We display the mass (M) and length (L) distributions of this catalog of Galactic filaments in Fig.\ref{fig:ML_phasespace}. Molecular filaments ($\sim 20\,000$; grey dots) cover more than eight orders of magnitude in mass ($M\sim 0.01-5\times10^6$~M$_\odot$), four orders of magnitude in length ($L\sim 0.03-300$~pc), and four orders of magnitude in terms of line mass ($m=M/L\sim 1-10^4$~M$_\odot$~pc$^{-1}$). These filaments are however not uniformly distributed across the mass-length (M-L) phase-space. Instead, most molecular filaments are primarily found along a main sequence in this plane indicating a physical correlation between the M and L of these objects. 

The origin of the observed distribution of filaments in the M-L parameter space is connected to the hierarchical structure of the molecular gas in the ISM and its fractal nature \citep[e.g.,][]{Larson1981,Elmegreen1996}. Cloud-scale filaments, usually studied at low-resolution, harbour parsec-size filaments that further break down into smaller filaments at sub-parsec scales when observed at  increasing resolution and sensitivity. The Orion A cloud appears as paradigmatic example of this nested gas organization. This $\sim100$~pc filamentary cloud \citep{Grossschedl2018} identified in large-scale surveys \citep{Lombardi2011,Nishimura2015} resolves into a series of $\sim 1-10$~pc-long filaments using dedicated molecular \citep{Nagahama1998} and continuum \citep{Johnstone1999,Schuller2021} observations which further fragment into even smaller filaments at sub-parsec scales when observed at interferometric resolutions \citep{Wiseman1998,Hacar2018,Monsch2018}. 
\citet{Hacar2022} identified a series of fundamental scaling relations describing this hierarchical organization. These scaling relations are analogous to the so-called Larson's relations \citep{Larson1981} associated to the inertial turbulent regime in molecular clouds. Compared to the original Larson's work, who assumed molecular clouds to be spherical (i.e. $2R=L$ or $A=1$), these new description is applied to elongated geometries in which the length $L$ and radius $R$ of filaments can vary independently.

First, \citet{Hacar2022} found that the total mass ($M$) and length ($L$) in most filamentary structures in the ISM (grey dots) follow a fundamental M-L relation such as 
\begin{equation}\label{eq:ML}
    L = a \cdot M^{0.5},
\end{equation}
where the slope $\alpha=0.5$ is set by the turbulent fragmentation of the gas at different scales (see also Sect.~\ref{sec:relevanttimes}).
As illustrated in Fig.\ref{fig:ML_phasespace}, Eq.~\ref{eq:ML} describes the global mass-length dependence among filaments at different scales (solid blue line)\footnote{Eq.~\ref{eq:ML} assumes a typical power-law dependence $\alpha=0.5$ describing the mean M-L relationship in filaments $L\propto M^\alpha$ varying between $\alpha=0.5\pm 0.2$, according to \citet{Hacar2022} (see their Eq.14)}. 

The correlation described by Eq.~\ref{eq:ML} for filaments is similar to the mass-size dependence observed in molecular clouds  \citep[$R\propto M^{0.5}$; ][]{Solomon1987} and is expected from the combination of the first ($\sigma\propto R^{0.38}$) and second ($\sigma\propto M^{0.20}$) Larson's relations \citep{Larson1981,Kauffmann2010a}. 
The filament M-L correlation is observed over almost three orders of magnitude in length in individual, high-dynamic range surveys such as Hi-GAL \citep{Schisano2020} and it is extended towards both larger and smaller scales by other dedicated Galactic Plane and interferometric surveys, respectively \citep[see ][for a full discussion]{Hacar2022}. 
Resolution and sensitivity limitations cause observations to selectively sample different regimes of this M-L main sequence. 
For instance, comparisons between overlapping surveys demonstrate how low-sensitivity, ground-based observations \citep[e.g., ATLASGAL; see][]{Schuller2009,Li2016} selectively sample filaments with larger masses at similar scales with respect to high-sensitivity, space-based studies \citep[i.e., {\it Herschel}; see Fig.~28 in][]{Schisano2020}. These massive filaments correspond to those targets with the highest surface densities (and thus observationally brightest) at a given scale that, although following the same M-L relation described by Eq.~\ref{eq:ML}, are found to be described by lower normalization values $a$.
These results suggest a potential systematic dependence of the M-L scaling relation with the gas surface density in analogy to the expected variations reported for the original Larson's relations \citep{Heyer2009,Camacho2023}.

Second, the M-L relation described by Eq.~\ref{eq:ML}  is accompanied by a similar the density-length (n-L) relation following \citep[see Fig.~4 in ][]{Hacar2022}\footnote{\citet{Hacar2022} report a n-L dependence such as $n\propto L^{\beta}$, with $\beta=-1.1\pm0.1$ (see their Eq.13). To simplify further analytic calculations we have approximated this exponent as $\beta=-1$.}
\begin{equation}\label{eq:nL}
    n \simeq b_0 \cdot L^{-1},
\end{equation}
with $b_0\sim 10^4$~cm$^{-2}$, and  
where
\begin{equation}\label{eq:n}
n =\frac{mass}{volume}=\frac{1}{\mu(H_2)}\frac{M}{\ \pi R^2 L}
\end{equation}
describes the average density of a filament derived in the reduced number of targets where simultaneous FWHM (=$2 R$), length, and mass measurements are reported in the literature.
The scaling relations described by Eqs.~\ref{eq:ML} and \ref{eq:nL} can be understood as a consequence of the self-similar nature of the gas organization in the ISM, where shorter and less massive filaments are representative of deeper levels in the hierarchy where the typical densities are higher.
Derived from a limited sample of filaments \citep[see Sect.~4 in][for a discussion]{Hacar2022}, we remark that the slope in this density scaling relation (Eq.~\ref{eq:nL}) is the most relevant source of uncertainty in our derivations. 

Third, molecular line observations show that the gas kinematics inside filaments obeys a velocity dispersion-size ($\sigma$-L) relation such as \citep[see Fig.~6 in ][]{Hacar2022} 
\begin{equation}\label{eq:sigmaL}
\sigma_{tot}=c_s\left(1+\frac{L}{0.5 \mathrm{pc}} \right)^{0.5}, 
\end{equation}
similar to the first Larson's scaling relation \citep{Larson1981}, where $\sigma_{tot}$ refers to the total velocity dispersion (i.e. thermal plus non-thermal motions) along the line-of-sight and $c_s=\sqrt{\frac{k T_\mathrm{K}}{\mu(H_2)}}$ denotes the gas sound speed with a gas kinetic temperature $T_\mathrm{K}$ with $k$ the Boltzmann constant. Its combination with the expected line-mass ($m=M/L$) for an isothermal filament in hydrostatic equilibrium \citep{Stodolkiewicz1963,Ostriker1964} defines the virial mass
\begin{equation}\label{eq:mvir}
    m_{vir}= \frac{2\sigma_{tot}}{G} = \frac{2 c_s}{G} \cdot \left(1+ \frac{L}{0.5 \mathrm{pc}} \right)^{0.5},
\end{equation}
with $G$ as the gravitational constant.
Eq.~\ref{eq:mvir} describes the state of energy equipartition in filaments (i.e. balance between gravitational and kinetic energy, $|E_g|/E_K=2$). While evolving over time, and not in equilibrium \citep{BallesterosParedes2006}, subvirial filaments ($m\lesssim m_{vir}$ and thus $|E_g|\lesssim 2\ E_K$) are expected to remain radially stable.
Instead, supervirial  filaments ($m> m_{vir}$, or $|E_g|>2E_K$) are unstable under gravity and will collapse into a spindle on timescales comparable to their free-fall time \citep[i.e. $\tau_{col}\sim \tau_{ff}$; see][]{Inutsuka1997}. We show the expected values for $m_{vir}$ derived from Eq.\ref{eq:mvir} at $T_\mathrm{K}=$~10~K in Fig.\ref{fig:ML_phasespace} (red solid line)\footnote{We note that particularly the lower end of the $m_{vir}$ function might shift to higher masses at higher $T_\mathrm{K}$ values.}.

\subsection{\bf HFS as precursors of young star clusters}\label{sec:HFSclusters}

In Fig.\ref{fig:ML_phasespace} we display the position of different HFS reported in the literature and compiled in Appendix~\ref{sec:appendixA} (see Table~\ref{table:HFS_data} and reference therein). 
The HFS included in our sample show typical masses of $M_{HFS}\sim 100-2\times10^4$~M$_\odot$ and sizes up to $L_{HFS}\sim 5$~pc.
The above observables translate into HFS line masses of $m_{HFS}=\frac{M_{HFS}}{L_{HFS}}\sim 200-5\times 10^3$~M$_\odot$~pc$^{-1}$ and average surface densities $\Sigma_{HFS}=\frac{M_{HFS}}{\pi (L_{HFS}/2)^2}=1.4\times10^{22}-3.1\times10^{23}$~cm$^{-2}$ as expected for these systems \citep{Myers2009a,Kumar2020}.

The HFS explored in our work populate a parameter space similar to the most massive gas clumps identified in different Galactic surveys \citep[e.g.,][]{Urquhart2014,Urquhart2018,Barnes2021,Elia2017,Elia2021_HiGALcat,Peretto2023}.
In particular, the properties of these HFS mimic those of the most massive 
ATLASGAL clumps \citep{Urquhart2014} with median values of $M_{AGAL}\sim 2\times 10^3\ M_\odot$, $R_{AGAL}\sim 0.95$~pc,  $\Sigma_{AGAL}\sim 4\times 10^{22}$~cm$^{-2}$, and $A_{AGAL}\sim 1.38$, respectively. 
Their distribution also roughly follows the same M-L relation (dashed black line) derived for the complete ATLASGAL clump sample \citep[light blue diamonds; ][]{Urquhart2018} as well as from other Galactic Plane clump surveys such as Hi-GAL \citep[not shown;][]{Elia2021_HiGALcat}. 
These similarities suggest that some of these apparently compact clumps might resolve into HFS when observed at high resolution \citep[e.g., see][for some recent examples]{Anderson2021}.
Likewise, most of the HFS compiled in our sample lie above the mass threshold for high-mass star-formation proposed in previous studies \citep[dotted purple line; see][]{KauffmannPillai2010} which denotes the potential of these regions to form massive stars. 

Different studies have proposed a direct connection between HFS and the origin of young star clusters \citep[e.g.,][]{Myers2009a,Peretto2013,Zhou2022,Kumar2020}.
Analogous to the ATLASGAL clumps \citep[see also][]{Urquhart2014,Urquhart2018}, our results postulate these HFS as potential gas precursors of typical Milky Way proto-clusters with embedded stellar masses between $M_\star\sim 50-2000$~M$_\odot$  (assuming a fiducial star formation efficiency of $\epsilon_{SF}\lesssim$~10\%, i.e. $M_\star\sim\epsilon_{SF}\cdot M_{HFS}$, in a first order approximation) and thus contain up to a few massive stars \citep{Kroupa2002}. We note however that different gas conditions may be needed to explain the origin of young massive clusters with $M_\star > 10^4$~M$_\odot$ \citep[see][for a discussion]{Longmore2014_PP6,Krumholz2020}.

Remarkably, these HFS (as well as most of the ATLASGAL and Hi-GAL clumps) depart from the scaling relation described by Eq.~\ref{eq:ML} and show systematically higher total (M) and line ($m$) masses than regular filaments on scales of $\sim 0.5-5$~pc populating the lower right side of the M-L diagram (blue crosses). 
Also, while most of the ISM filaments can be classified as as subvirial ($m<m_{vir}$), all HFS appear to be supervirial ($m>m_{vir}$, beneath the red line) in opposition to filaments with similar masses.

\subsection{A toy model to explore the M-L parameter space}\label{sec:AR_col}

\begin{figure*}[ht!]
\centering
\includegraphics[width=1.\textwidth]{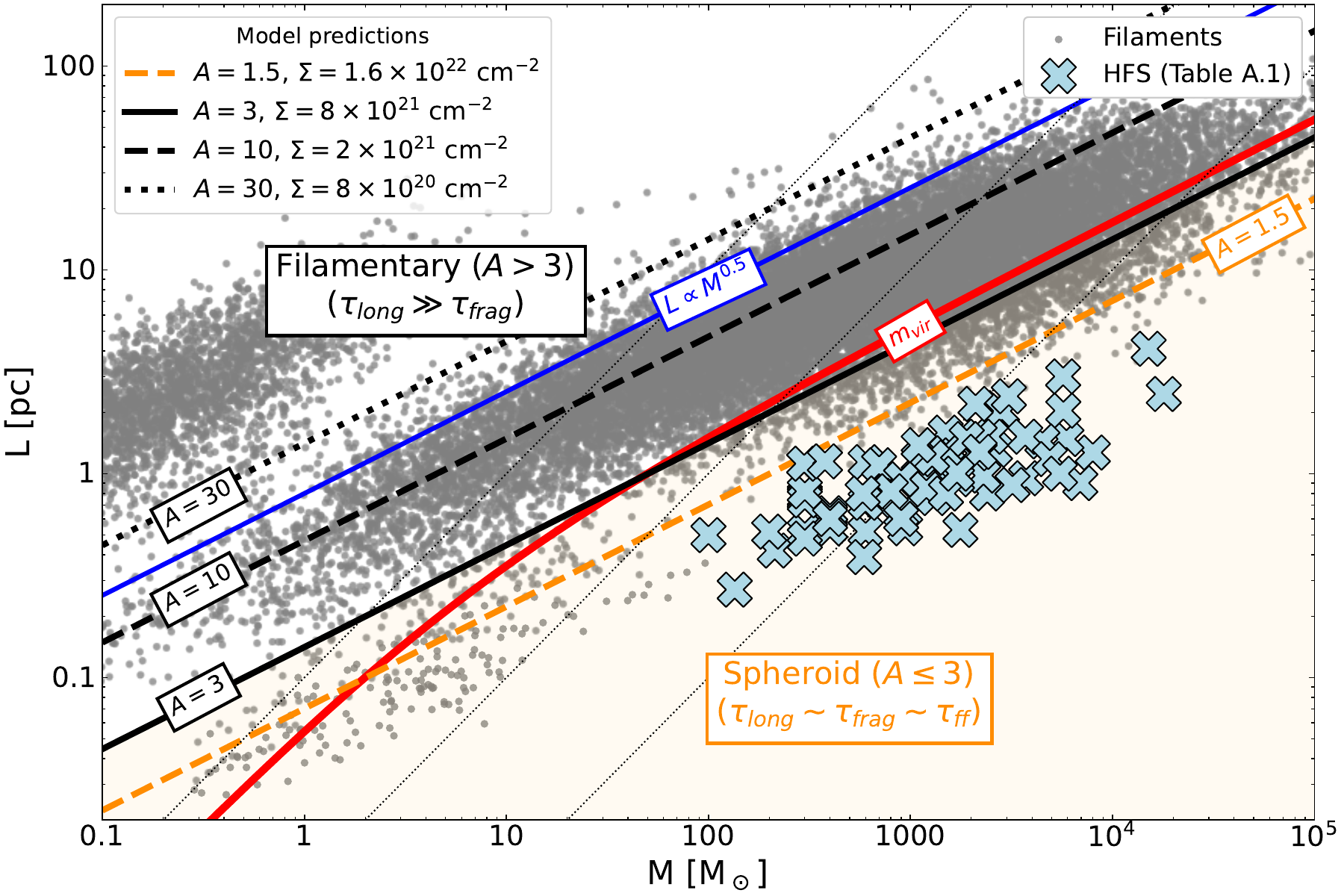}
      \caption{
      Model predictions (Sect.~\ref{sec:AR_col}) compared to the location of the ISM filaments in the M-L phase-space (see also Fig.~\ref{fig:ML_phasespace}). For simplicity, we display only those molecular filaments.
      Using Eq.\ref{eq:a_AR}, different lines indicate the expected normalization values for characteristic filamentary ($A$~=~30, 10, and 3; black lines) and prolate spheroids ($A<3$, orange shaded area) geometries. We added also the typical $A$ observed for HFS \citep[$A\leq 1.5$; orange dashed line; e.g.,][]{Kumar2020} (see also legend for their corresponding surface densities according to Eq.\ref{eq:a_Sigma}).
      }
\label{fig:ML_AR}
\end{figure*}

We aim to understand and characterize the different locations of filaments and HFS throughout the M-L parameter space.
For that, we create a toy model describing a generalized form of Eq.~\ref{eq:ML} with different normalization values $a$. As suggested by \citet{Hacar2022}, this normalization value is likely connected to the mean column density of these filaments since denser filaments appear to be on the lower side of this relation.
To explore this hypothesis we can estimated the (average) surface density of a filament as
\begin{equation}\label{eq:surfden}
    \Sigma = \frac{mass}{area}=\frac{\pi R^2 L n}{2RL}=\frac{\pi}{2}Rn.
\end{equation}
For convenience, and similar to \citet{Pon2012}, we also defined the aspect ratio of a filament $A=L/FWHM$ as function of its radius ($R=FWHM/2$) as
\begin{equation}\label{eq:AR} 
A=\frac{L}{2R} \rightarrow R=\frac{1}{2}\frac{L}{A}.
\end{equation}
Inserting Eqs.\ref{eq:nL} and \ref{eq:AR} in Eq.\ref{eq:surfden}, it follows that
\begin{equation}\label{eq:surfden_AR}
    \Sigma =\frac{\pi}{2}Rn = \frac{\pi b_0}{4}\frac{1}{A}\approx \frac{2.3\times 10^{22} }{A}\ \mathrm{cm}^{-2},
\end{equation}
which links the observed (average) surface density of filaments with their (typical) aspect ratio. 

Coming back to the M-L relation, we can replace Eq.~\ref{eq:ML} assuming a constant  density coefficient $b_0$ in Eq.~\ref{eq:nL} such as
\begin{equation}
    L = a \cdot M^{0.5} = a (\pi R^2 L n)^{0.5}=a (\pi b_0)^{0.5}\cdot R
\end{equation}
and combine it with Eq.~\ref{eq:AR} to express the normalization parameter $a$ by means of aspect ratio {\bf$A$} as
\begin{equation}\label{eq:a_AR}
    a= \frac{2A}{(\pi b_0)^{0.5}}.
\end{equation}
Alternatively, by using Eq.~\ref{eq:surfden_AR}, this parameter can also be written as
\begin{equation}\label{eq:a_Sigma}
    a= \frac{(\pi b_0)^{0.5}}{2}\frac{1}{\Sigma}.
\end{equation}
Equations \ref{eq:a_AR} and \ref{eq:a_Sigma} connect the normalization of the M-L relation in filaments with their (typical) aspect ratio $A$ and average surface density $\Sigma$, respectively (an alternative derivation of these equations can be found in Appendix~\ref{ap:alternative}). In particular, filaments with smaller $A$, and therefore higher $\Sigma$, are expected to follow a M-L relation with a lower normalization value, while filaments with a longer $A$, and consequently lower $\Sigma$, are predicted to exhibit higher normalization values, which cause them to lie higher up in the M-L plane. Spheroidal clouds can be described as the asymptotic limit of filaments when $A\rightarrow$1. 
Starting from Eq.~\ref{eq:surfden}, it is trivial to demonstrated that the classical assumption of spherical symmetry (i.e. $2R=L$) leads to clouds with roughly constant $\Sigma$  \citep[e.g.,][]{Larson1981}.
Instead, 
different surface density values are permitted in filaments with varying $A>1$ according to Eq.~\ref{eq:surfden_AR}, where spheres appear in the asymptotic limit when $A\rightarrow$1.

Combined with Eq.~\ref{eq:ML}, Eq.~\ref{eq:a_Sigma} also leads to an additional and direct consequence of our model: the average column density in filaments should scale with their mass and length such as
\begin{equation}\label{eq:predict}
    \Sigma \propto \frac{M^{0.5}}{L},
\end{equation}
which is different from the expected correlation for a uniform sphere $\Sigma \propto \frac{M}{R^2}$. The correlation described by Eq.~\ref{eq:predict} may be intuitively clear when comparing filaments with the same mass M, where shorter filaments are expected to show higher surface densities. Likewise, filaments of the same length and increasing mass should correspond to higher surface densities. 

It is worth emphasizing that the above derivations assume a constant density coefficient $b_0$ and a variable normalization $a$ parameter (see also Appendix~\ref{ap:alternative}). Inverting these assumptions, that is having a variable $b_0$ and constant $a$, would lead into a direct proportionality between the filament surface density and aspect ratio such as $\Sigma \propto A$ following the same Eqs.~\ref{eq:ML} and \ref{eq:surfden}. This possibility seems to be ruled out given the anti-correlation seen between $\Sigma$ and $L$ in Fig.~\ref{fig:HGBS}, which closely follows the predictions of Eq.~\ref{eq:predict} (see Sect.~\ref{sec:observations} for a full discussion).

Equations \ref{eq:ML}-\ref{eq:a_Sigma} provide an idealized framework to explore the correspondence between the observed mass and length of filaments and HFS and additional physical parameters such as $A$ and $\Sigma$.
While oversimplified by the use of symmetries and average properties, the next sections will probe this toy model as a useful tool to describe the different physical regimes of the filamentary ISM.

\subsection{Comparison with observations}\label{sec:observations}

\begin{figure}[]
\centering
\includegraphics[width=1.\linewidth]{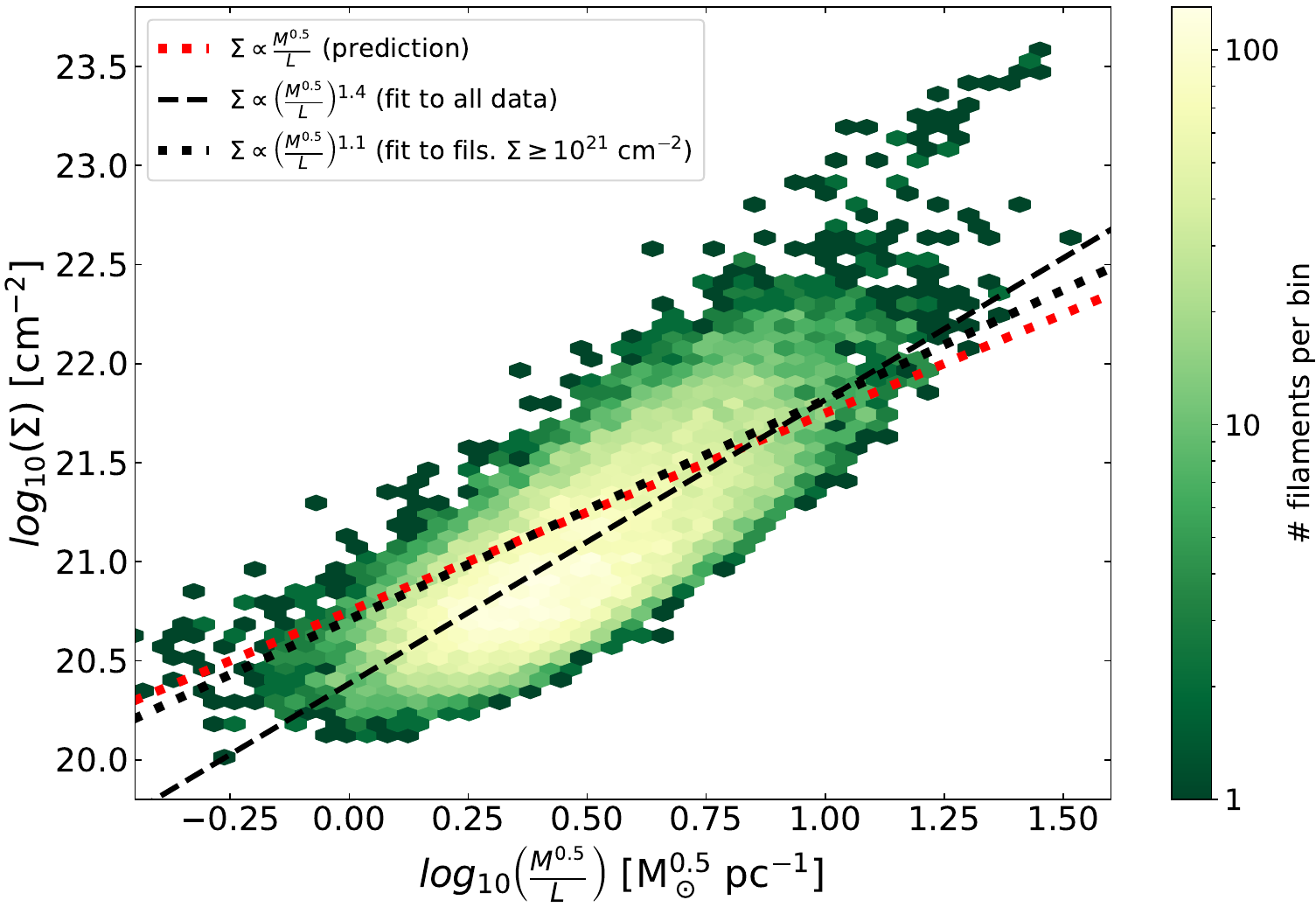}
      \caption{Comparison between the average column density $\Sigma$ and the $M^{0.5}/L$ ratio observed in the Hi-GAL filaments \citep[$>18\,000$ filaments;][]{Schisano2020} displayed using a hexagonal 2D density plot (see color bar).
      Overplotted on these data, we display the prediction from Eq.~\ref{eq:predict} (dotted red line) as well as different fits to the data (dashed and dotted black lines, see legend).
      }
\label{fig:HGBS}
\end{figure}

\begin{figure*}[ht!]
\centering
\includegraphics[width=1.\linewidth]{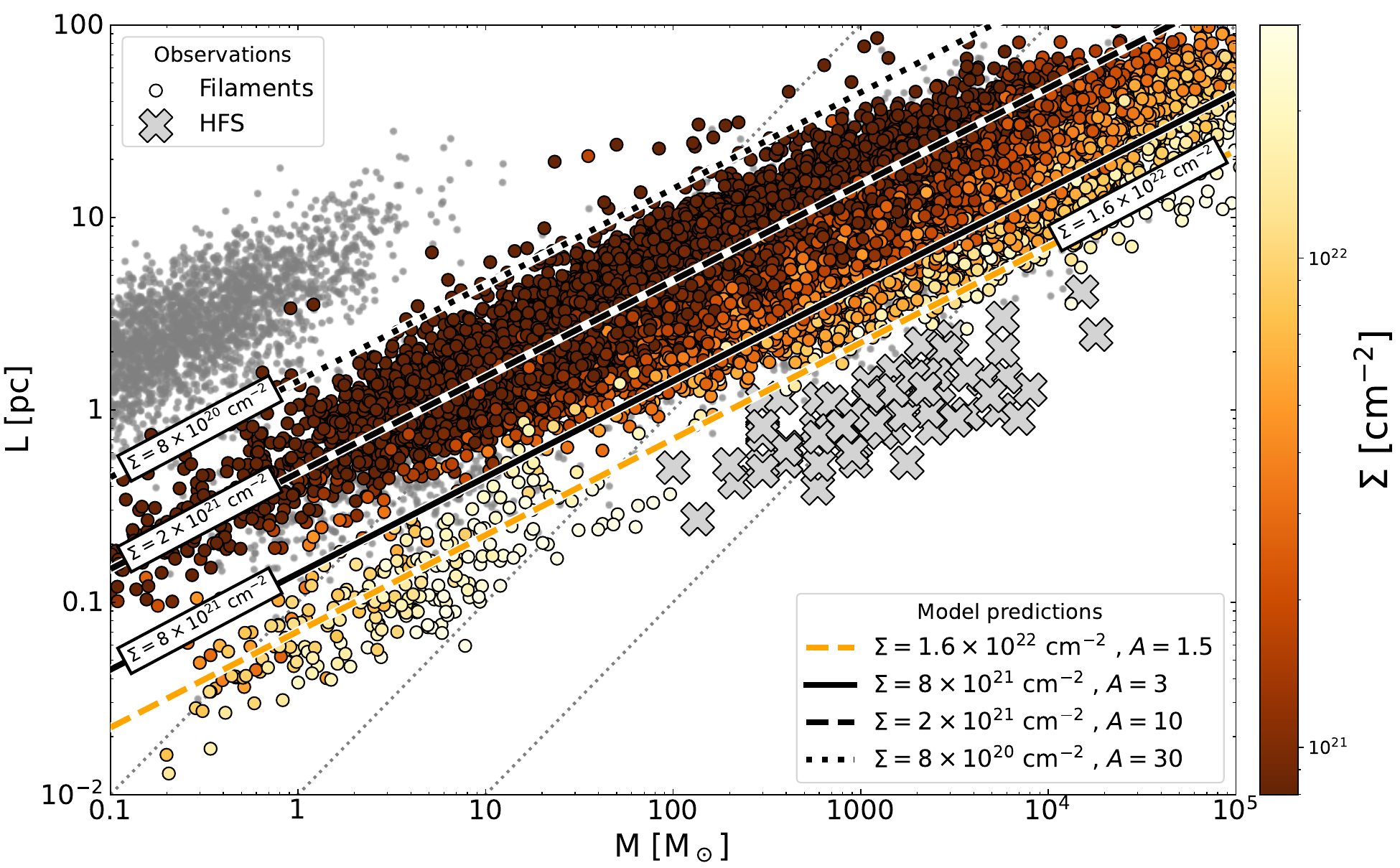}
      \caption{Mass-Length distribution of molecular filaments (circles) similar to Fig.~\ref{fig:ML_phasespace}, this time color-coded by their average column density $\Sigma$ (see color bar). The majority of these $\Sigma$ measurements ($>18,000$) were obtained by the Hi-GAL survey \citep{Schisano2020} while few additional datapoints (152) are taken from our Paper III \citep{SocciPaperIII}.
      Analogous to Fig.~\ref{fig:ML_AR}, different lines indicate the predicted distribution of filaments with mean column density $\Sigma = [0.8, 2, 8, 16]\cdot 10^{21}$~cm$^{-2}$ (see legend) according to Eqs.~\ref{eq:ML} and \ref{eq:a_Sigma}, corresponding with aspect ratios of $A=[30, 10, 3, 1.5]$, respectively, following Eq.~\ref{eq:a_AR}. 
      The location of all HFS (grey crosses) are indicated in this plot for comparison. We note that the position of the HFS would correspond with the expected location for prolate ($A\lesssim 1.5$) and high-column density ($\Sigma > 10^{22}$~cm$^{-2}$) structures according to the predictions of our model. 
      See also Fig.~\ref{fig:prediction_N_3panels}.
      }
\label{fig:prediction_N}
\end{figure*}

In Fig.\ref{fig:ML_AR} we show the corresponding M-L scaling relations expected for aspect ratios $A$ = 3, 10, and 30 (black lines) describing typical values for filamentary geometries obtained combining Eqs.\ref{eq:ML} and \ref{eq:a_AR}. We also show the corresponding line for $A$~=~1.5 (dashed orange line) characteristic of prolate objects such as HFS \citep[e.g.,][]{Kumar2020}.
According to these model predictions, standard filaments should present mass and length values corresponding with $A>$~3 and (average) surface densities of $\Sigma\lesssim 8\times10^{21}$~cm$^{-2}$ (see Eq.~\ref{eq:surfden_AR}). 
Previous observations reported molecular filaments show typical $A=5-15$ and maximum values of $A\sim$~30, with corresponding peak surface densities between $\Sigma_0=8\times10^{20}-8\times10^{21}$~cm$^{-2}$ \citep[e.g.,][]{Arzoumanian2019,Schisano2020}, in close agreement to our model\footnote{We remark that an $A\geq 3$ is usually employed as threshold for identifying gas structures in most filament-finding algorithms \citep[e.g., see][]{Arzoumanian2019}.}.

In Fig.~\ref{fig:HGBS}, we directly compare the prediction of Eq.~\ref{eq:predict} with the filament properties obtained by \citet{Schisano2020} from the analysis of the {\it Herschel} Hi-GAL survey \citep{Molinari2010} which obtained a census of $>18\, 000$ filaments across the Galactic Plane.
For each filament in this survey, we display the filament average column density $\Sigma$ and compare it with their corresponding $M^{0.5}/L$ ratio. Figure~\ref{fig:HGBS} shows a clear correlation between these observables along more than an order of magnitude in each axis.
A linear fit to the Hi-GAL filaments (in log-log space) leads to $\Sigma \propto \left(\frac{M^{0.5}}{L}\right)^{1.4}$ for all datapoints (dashed black line) and shows a good agreement (within the noise) with the expected $\Sigma \propto M^{0.5}/L$ dependence of our model (dotted red line). 
A better fit, with $\Sigma \propto \left(\frac{M^{0.5}}{L}\right)^{1.1}$, is obtained if we only consider those filaments
with $\Sigma>10^{21}$~cm$^{-2}$ where the Hi-GAL sample is expected to be less affected by completeness \citep[dotted black line; see ][for a discussion of this threshold]{Schisano2020}.
A roughly similar dependence (with $\Sigma \propto \frac{M}{L}$) has been reported by \citet{Schisano2020} (see their Fig.~26) for these Hi-GAL filaments as well as by \citet{Arzoumanian2019} (see their Fig.~10, panel a) in the case of the {\it Herschel} Gould Belt Survey (HGBS) in nearby clouds.

As additional consequence of Eq.~\ref{eq:predict}, our model predicts a stratified distribution of the ISM filaments in the M-L phase space.
In particular, molecular filaments are expected to populate the M-L plane following a similar functional dependence ($L=a\cdot M^{0.5}$; Eq.~\ref{eq:ML}) but with different normalization values $a$ depending on their mean column densities, i.e. $a\propto 1/\Sigma$ (Eq.~\ref{eq:a_Sigma}).  
We explored these predictions in Fig.~\ref{fig:prediction_N} by displaying again the M-L distribution of ISM filaments this time (and when available) color-coding each target by its corresponding average column density $\Sigma$ (see colorbar). Gaps in this diagram correspond to different observational biases and incompleteness affecting our filament sample \citep[see][for a discussion]{Hacar2022}.
A systematic variation is seen as function of $\Sigma$ across this plot, where filaments with similar column densities define parallel bands of different colors, and where filaments with larger column densities are preferentially located towards the lower side of this distribution (see also Appendix~\ref{sec:appendixB}).
Including the unavoidable data noise, the column density values predicted by our model are in good agreement with the observed filament column densities across the entire M-L space.

\begin{figure*}[h!]
\centering
\includegraphics[width=0.85\linewidth]{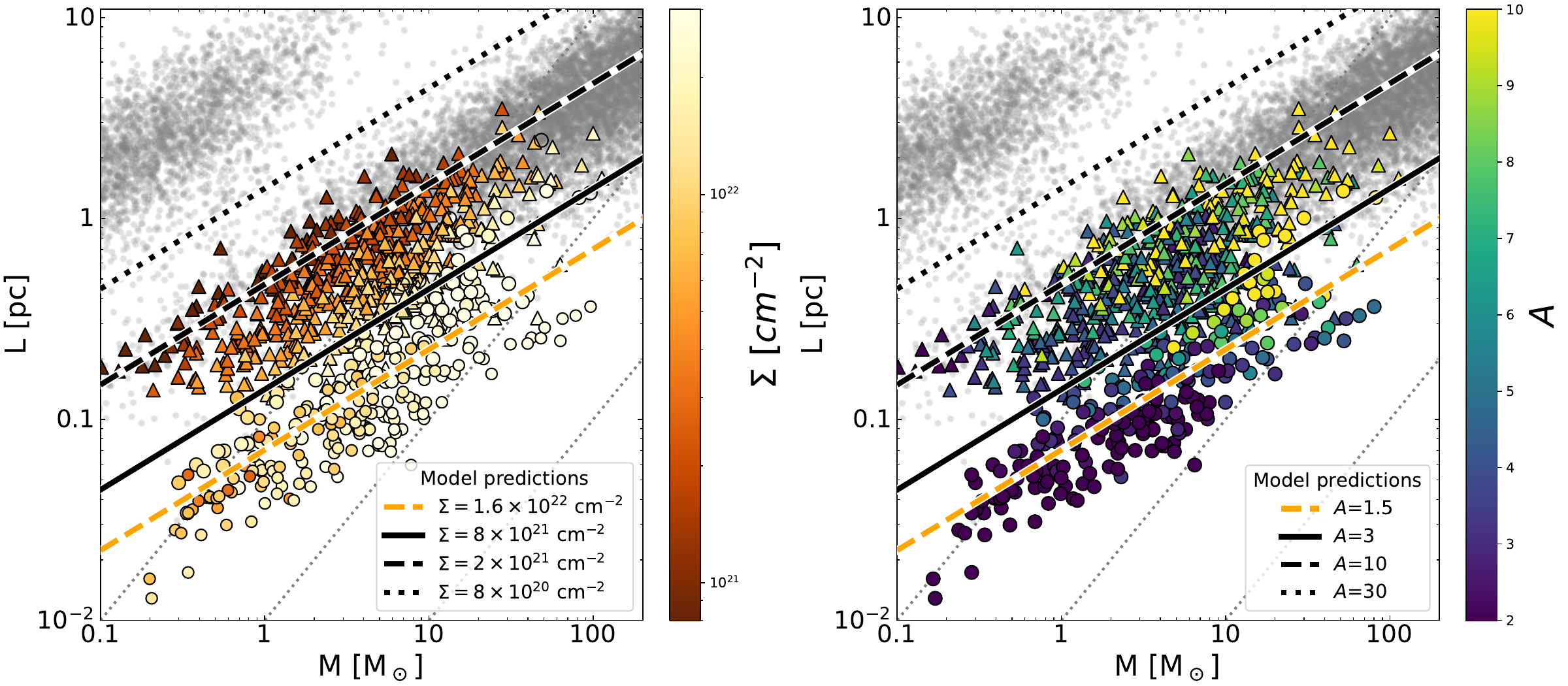}
      \caption{Mass-Length distribution of the nearby
      HGBS filaments \citep[triangles;][; priv. comm.]{Arzoumanian2019} and fibers \citep[circles][]{Hacar2022,SocciPaperIII,SocciPaperIV} color-coded by 
      their mean column density $\Sigma$ (left panel) and aspect ratio $A$ (right panel). 
      Labels and lines are similar to Fig.~\ref{fig:prediction_N}.
      }
\label{fig:prediction_N_HGBS}
\end{figure*}

Likewise, and according to Eqs.~\ref{eq:surfden_AR} and \ref{eq:a_AR}, the filament aspect ratio and mean column density should be anti-correlated $A\propto \Sigma^{-1}$ leading to a scaling relation such as $L\propto A\cdot M^{0.5}$.
Lines of constant $A$ at different $L$ values would suggest that the filament FWHM is therefore not constant but can vary significantly when comparing filaments at different scales ($L\propto A\cdot M^{0.5} \rightarrow FWHM\propto M^{0.5}$).
We have tested this hypothesis in Fig.~\ref{fig:prediction_N_HGBS} by comparing the M-L distribution of those resolved filaments identified by the HGBS \citep[solid triangles; ][; priv. communication]{Arzoumanian2019} as well as in different fiber surveys \citep[solid circles; ][]{Hacar2013,Hacar2017b,SocciPaperIII} and show them color-coded by their mean column density $\Sigma$ (left panel) and aspect ratio $A$ (right panel), respectively. Inversely proportional to the previously discussed dependence with $\Sigma$ (left panel; similar to Fig.~\ref{fig:prediction_N}), these resolved filaments show a tentative stratification as function of $A$ (right panel) with filaments with lower aspect ratios populating the lower part of these diagrams roughly following our model predictions (see legend and color bar). While undoubtedly less systematic, we note that observational, selection, and algorithmic biases between these surveys \citep[see][for relevant discussions]{Andre2019,Hacar2022} make these $A$ estimates intrinsically noisier than direct $\Sigma$ measurements.
Unfortunately, these comparisons cannot be extended to the Hi-GAL sample given the limited resolution of {\it Herschel} to resolve most Galactic filaments at kpc distances. 
Additional high-resolution data are needed to confirm this predicted $FWHM\propto M^{0.5}$ dependence\footnote{We note that this $FWHM\propto M^{0.5}$ dependence is equivalent to the expected $FWHM\propto L$ proposed by \citet{Hacar2022} given the M-L scaling in Eq.~\ref{eq:ML}.
See also \citet[][]{Panopoulou2022,Andre2022Gina} for a discussion.}.

Continuing the above trend, the predicted values from Eqs.\ref{eq:a_AR} for $A=$~3 separate most of the ISM filaments ($A>$~3)\footnote{Our M-L diagram actually displays the observed filament lengths $L_{obs}$ reported in the literature. De-projection effects (i.e. $L_{obs}=L\cdot cos \theta$) would therefore move most filaments upwards by an average factor $\frac{1}{<cos\ \theta>}$, where $<cos\ \theta>|_0^{\pi/2}=\frac{1}{\pi/2}\cdot\int_0^{\pi/2}cos\ \theta \ d\theta=\frac{2}{\pi}$.} from the location of the most massive and compact HFS ($A<$~3) in Fig.~\ref{fig:prediction_N}. This transition at $A=3$ corresponds to regions with $\Sigma = 8\times 10^{21}$~cm$^{-2}$ (Eq.\ref{eq:surfden_AR}; solid black line) and an average density $n>10^4$~cm$^{-3}$ at scales of $\sim$~1~pc (Eq.\ref{eq:nL}). The location of the observed HFS coincides with the expected parameter space for spheroids with low aspect ratios $A\leq 1.5$ and high column densities $\Sigma\geq 2\times 10^{22}$~cm$^{-2}$ (dashed orange line), in close agreement with observational results \citep{Myers2009a,Kumar2020,Morii2023}.
Our findings suggest a change of the gas geometry, between filamentary to spheroidal, during the formation of these compact HFS as driver for their formation.

\section{Evolutionary timescales and accretion rates}\label{sec:timescales}

\subsection{Relevant timescales}\label{sec:relevanttimes}

The evolution of radially stable filamentary structures (i.e. sub-virial or $m\lesssim m_{vir}$) in the M-L plane can be described by the combined effect
of three independent processes, each with a specific signature in this parameter space (see top left corner in Fig.~\ref{fig:ML_cartoon}): (a) {\it fragmentation} (splitting an object into pieces of smaller M and L, and thus moving them diagonally downwards in the M-L space); (b) {\it accretion} (moving objects horizontally towards higher M at constant L); and (c) {\it longitudinal collapse} (moving objects downwards in L at constant M).
We can characterize the relative importance of these three independent processes by comparing their corresponding timescales \citep[see][for a full discussion]{Hacar2022}.

\begin{figure}[t]
\centering
\includegraphics[width=1.\linewidth]{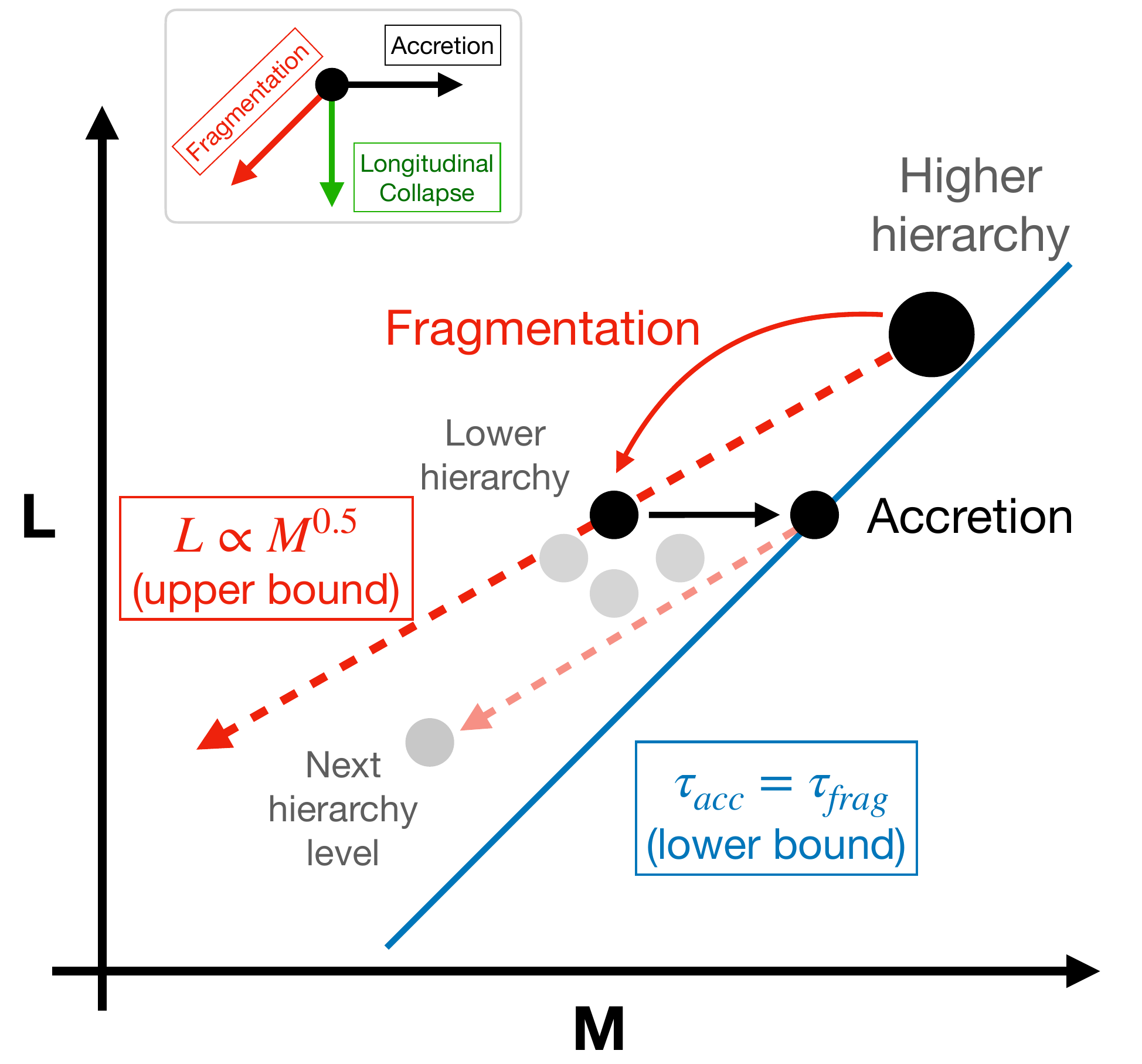}
      \caption{Schematic description of the evolution of sub-critical filaments in the M-L phase-space experiencing hierarchical fragmentation. See text for a description.}
\label{fig:ML_cartoon}
\end{figure}

\begin{figure*}[ht!]
\centering
\includegraphics[width=1.\textwidth]{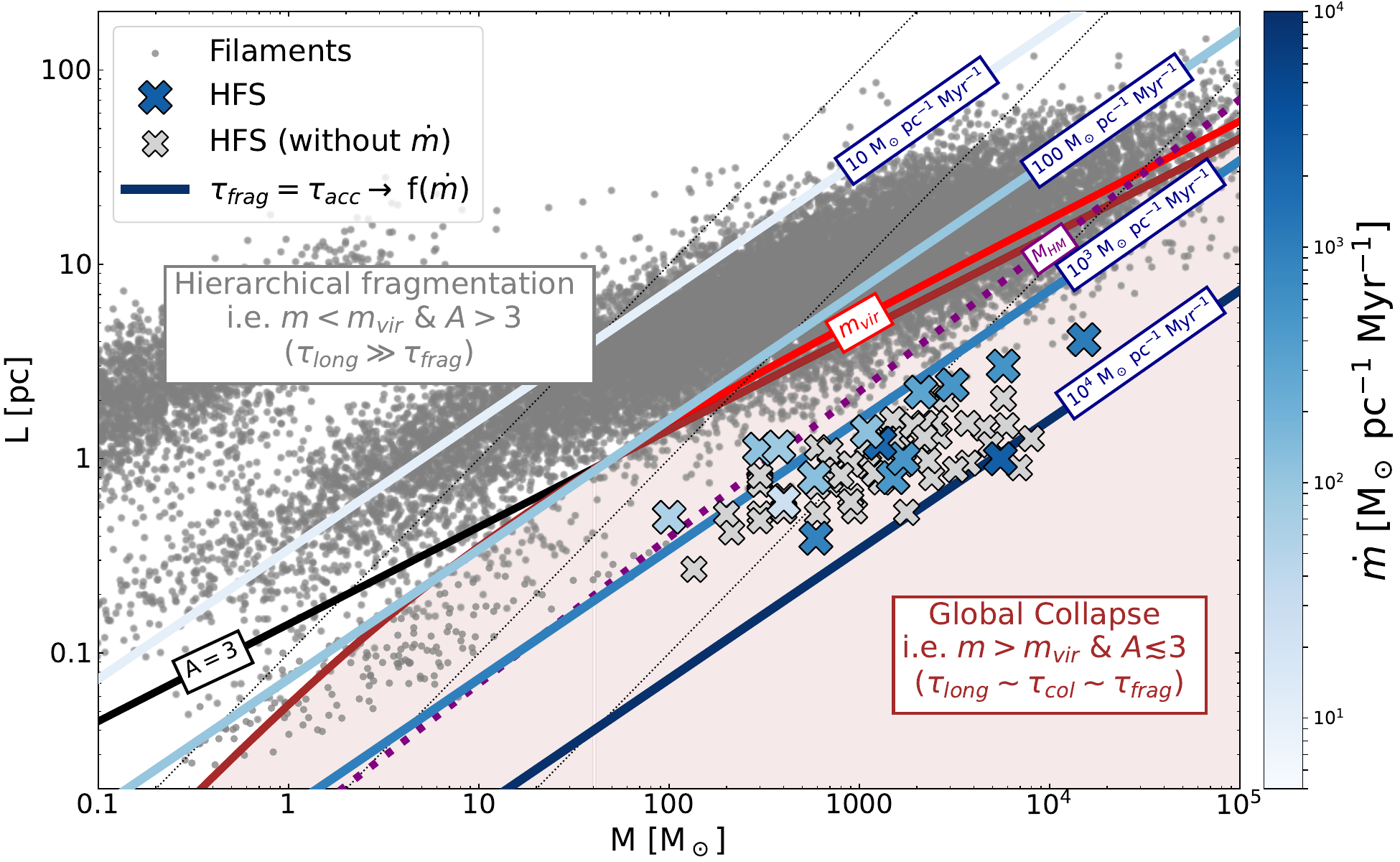}
      \caption{Mass-Length plot similar to Fig.~\ref{fig:ML_phasespace}, this time showing curves where $\tau_{frag}=\tau_{acc}$ for different accretion rates, namely  $\dot{m}=[10,100,10^3,10^4]$~M$_\odot$pc$^{-1}$Myr$^{-1}$ (straight blue lines), using different colors (see individual labels and color bar).
      We display those HFS with accretion measurements using the same color-code (see color bar on the right). 
      We highlight the region with expected global (radial + longitudinal) collapse (shaded brown area; where $\tau_{long}\sim\tau_{ff}\sim\tau_{frag}$) corresponding to the area where both spheroidal ($A\leq 3$; black line) and supervirial ($m>m_{vir}$; red curve) conditions are satisfied (see also Fig.~\ref{fig:ML_AR}). 
      }
\label{fig:ML_mdot}
\end{figure*}

First, the fragmentation timescale ($\tau_{frag}$) is defined as \citep{Larson1985,Inutsuka1992}
\begin{equation}\label{eq:tau_frag}
    \tau_{frag}= \frac{3}{2\sqrt{\pi G n}} = 1.66 \left( \frac{n}{10^3 \mathrm{cm}^{-3}} \right) ^{-0.5} \ \mathrm{Myr}
\end{equation}
which, applying Eq.\ref{eq:nL}, can be re-written as\footnote{We note that \citet{Hacar2022} derives a similar correlation but with a slightly different power dependence, that is $\tau_{frag}= 0.5 \left(\frac{L}{\mathrm{pc}}\right)^{0.55}$, by using the original $n\propto L^{-1.1}$ relation rather than its simplified version $n\propto L^{-1}$ used in Eq.\ref{eq:nL}. This small changes has no influence on the final results while simplifying all calculations.}
\begin{equation}\label{eq:tau_frag2}
    \tau_{frag}= 0.5 \left(\frac{L}{\mathrm{pc}}\right)^{0.5} \ \ \mathrm{Myr}.
\end{equation}

Second, the accretion timescale ($\tau_{acc}$) is defined as the time to double the mass of a filament given an accretion rate $\dot{m}$ (in units of M$_\odot$~pc$^{-1}$~Myr$^{-1}$):
\begin{equation}\label{eq:tau_acc}
    \tau_{acc}= \frac{M/L}{\dot{m}} \ \mathrm{Myr}.
\end{equation}
A connection between the filament line mass $m=M/L$ and the gas accretion rate $\dot{m}$ may be expected if accretion would be gravitationally driven \citep[e.g.,][]{Heitsch2013a}, although we remark that this condition is not imposed in our model.
Instead, our generic definition of $\tau_{acc}$ aims to describe the timescale in which accretion has a noticeable influence in the filament gas dynamics irrespective of its nature \citep[e.g., turbulent accretion; see][]{Padoan2020}.

Third, the timescale for longitudinal collapse ($\tau_{long}$), or the time for collapse along the main filament axis, can be described as a function of the free-fall time ($\tau_{ff}=\sqrt{\frac{3\pi}{32 G n}}$) like \citep{Pon2012,Toala2012,Hoemann2023}
\begin{equation}\label{eq:tau_long}
    \tau_{long}= \frac{\sqrt{32\cdot A}}{\pi}\tau_{ff}=1.9\cdot \sqrt{A} \left(\frac{n}{10^3 \mathrm{cm}^{-3}}\right)^{-0.5} \  \mathrm{Myr}.
\end{equation}
The combination of Eqs.\ref{eq:tau_frag} and \ref{eq:tau_long} to eliminate the density dependence gives
\begin{equation}\label{eq:tau_long_vs_frag}
    \tau_{long}= 1.14\sqrt{A} \cdot \tau_{frag},
\end{equation}
which demonstrates that the time for longitudinal collapse $\tau_{long}$ of filaments ($A>3$) is longer than the fragmentation timescale (i.e. $\tau_{long}\gg \tau_{frag}$) and therefore is dynamically subdominant in elongated geometries  \citep[see also][]{Pon2012,Toala2012}.

\subsection{Low accretion rates in filaments}\label{sec:accretion_filaments}

Following Sect.~\ref{sec:relevanttimes}, the distribution of filaments in the M-L phase-space is determined by the competition between fragmentation and accretion processes ($A>3 \rightarrow \tau_{long}\gg\tau_{frag}\sim \tau_{acc}$). We illustrate these results in Fig.~\ref{fig:ML_cartoon} \citep[see also][for a full discussion]{Hacar2022}.
At any point of its hierarchy, a filament of mass M and length L (higher hierarchy level) will fragment into several sub-structures of lower M and shorter L (lower hierarchy level) in a timescale $\tau_{frag}$ (Eq.~\ref{eq:tau_frag2}). Driven by a fast turbulent fragmentation, the connection between two hierarchical levels is expected to statistically follow a random-walk realization such as $L\propto M^{0.5}$ (similar to Eq.~\ref{eq:ML}) with a normalization value determined by the higher level of this hierarchy (dashed red line).
Simultaneously, each sub-structure is expected to accrete mass from their surroundings at a rate given by their local $\dot{m}$ increasing its mass by a factor of two in a timescale $\tau_{acc}$ (Eq.~\ref{eq:tau_acc}). This mass gain (black arrow) can continue a maximum timescale comparable to $\tau_{acc}\geq \tau_{frag}$ (solid blue line) in which this structure is expected to fragment again into a lower hierarchy level. This balance can be estimated equating Eqs.\ref{eq:tau_frag2} and \ref{eq:tau_acc}:
\begin{equation}\label{eq:tau_acc_vs_frag}
    \tau_{acc}= \tau_{frag} \rightarrow L = 
    \left(\frac{2 M}{\mathrm{M}_\odot}\right)^{2/3}
    \left( \frac{\dot{m}}
    {\mathrm{M}_\odot\ \mathrm{pc}^{-1} \ \mathrm{Myr}^{-1}}\right) ^{-2/3} \mathrm{pc}
\end{equation}
Our model therefore predicts the location of hierarchical filaments in the M-L phase-space to be bracketed between the correlations defined by Eq.~\ref{eq:ML} (upper bound) and Eq.~\ref{eq:tau_acc_vs_frag} (lower bound).

Eq.\ref{eq:tau_acc_vs_frag} defines a power-law relationship in the M-L diagram (see Fig.~\ref{fig:ML_cartoon}), crossing this plot from its lower left to the upper right, with lower normalization values for higher accretion rates $\dot{m}$.
This can be seen after re-arranging Eq.~\ref{eq:tau_acc_vs_frag} like 
\begin{equation}\label{eq:mdot_pred}
    \dot{m}=\frac{2 M}{L^{3/2}}=\frac{2}{L^{1/2}}\cdot m,
\end{equation}
showing how $M$ and $m$ appear as good proxies of $\dot{m}$ according to our model. 
A similar correlation between $\dot{m}$ and $m$ has been recently found in simulations which suggests that the line mass of filaments is largely determined by their accretion rates \citep{Feng2024,Zhao2024_Pudritz}.

We display some characteristic values of $\dot{m}$ predicted using Eq.\ref{eq:tau_acc_vs_frag} in the M-L diagram in Fig.~\ref{fig:ML_mdot} (solid lines; see also color bar). In agreement to our model predictions, observations in nearby filaments report accretion rates $\dot{m}\sim 10-30$~M$_\odot$~pc$^{-1}$~Myr$^{-1}$ \citep[e.g.,][]{Palmeirim2013,Bonne2020}, with higher accretion rates $\dot{m}\leq 100$~M$_\odot$~pc$^{-1}$~Myr$^{-1}$ typically associated to star-forming filaments according to Galactic Plane surveys \citep{Schisano2014}. 
Our model satisfactorily reproduces the specific M and L distributions of hierarchical filaments in regions such as Musca, Taurus, and Orion \citep[see Fig.~9 in][]{Hacar2022}. 
The relatively low range of accretion rates $\dot{m}\sim 10- 100$~M$_\odot$~pc$^{-1}$~Myr$^{-1}$ found in filaments explains how these structures show a similar and continuous M-L relation across several orders of magnitude in mass and length.

\subsection{High accretion rates in HFS}\label{sec:accretion}

\begin{figure}[ht!]
\centering
\includegraphics[width=0.9\linewidth]{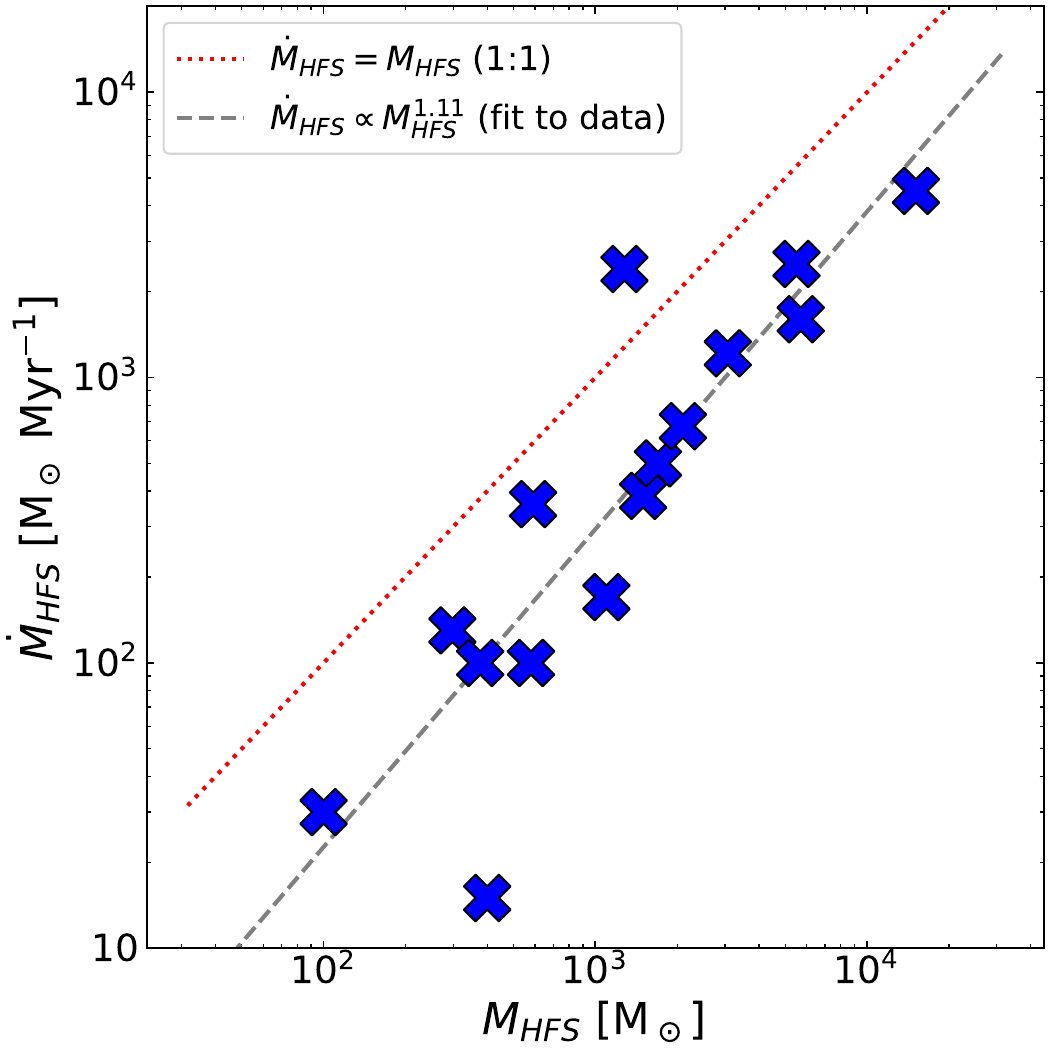}\\
\includegraphics[width=0.9\linewidth]{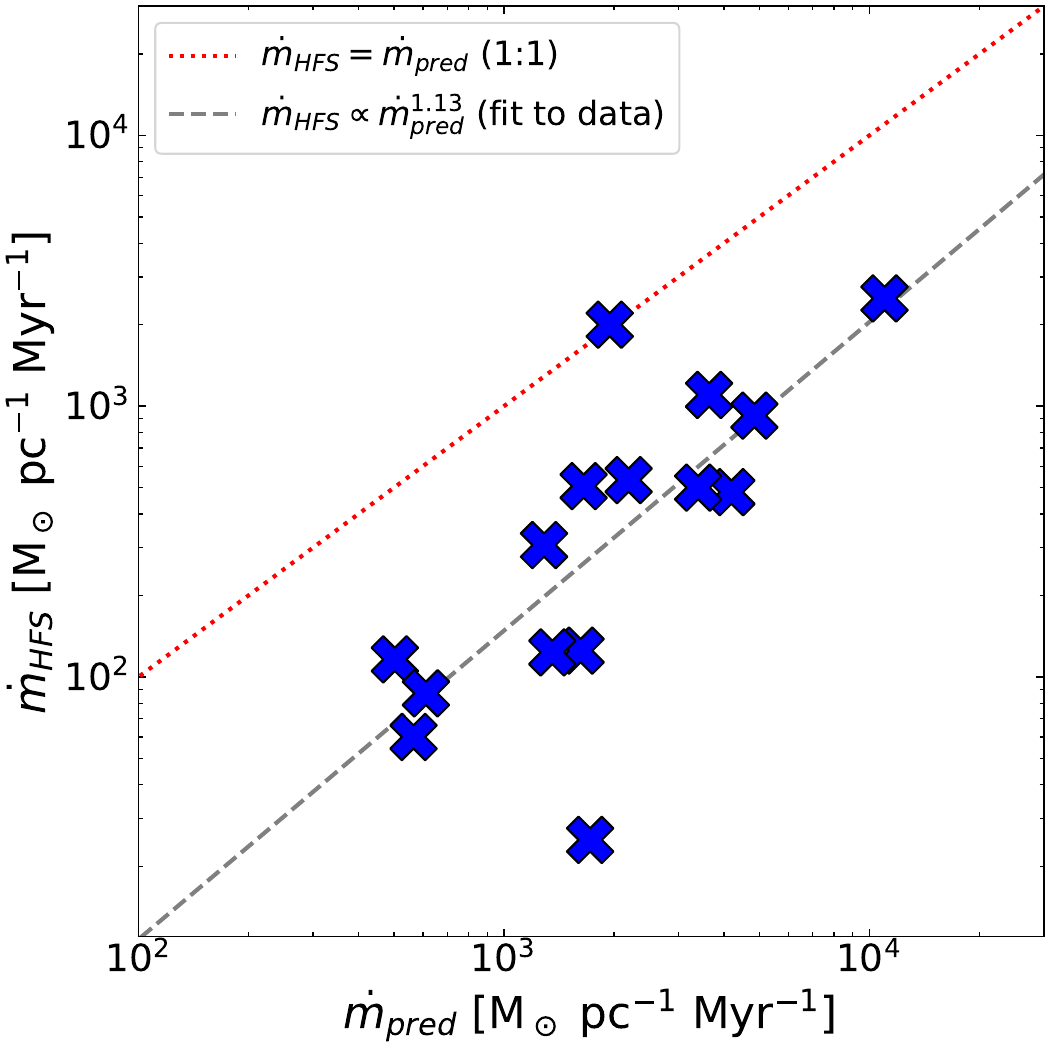}
      \caption{Accretion rates in HFS. 
      {\bf (Upper panel)}
      Comparison between observed total accretion rates ($\dot{M}_{HFS}$) and the HFS mass ($M_{HFS}$)
      reported in the literature for HFS (see also Table~\ref{table:HFS_data}).
      {\bf (Lower panel)}
      Comparison between observed accretion rates per-unit-length ($\dot{m}_{HFS}$) and those predicted following Eq.~\ref{eq:mdot_pred} ($\dot{m}_{pred}$). A red dotted line shows the expected 1:1 correlation between these values.
      A grey dashed line indicates a fit to the data (in log-log space) in both panels (see legend).
      }
\label{fig:HFS_mdot}
\end{figure}

The different locations of filaments and HFS in the M-L phase-space translate into different accretion rates $\dot{m}$ according to Eq.~\ref{eq:tau_acc_vs_frag}.
As seen in Fig.~\ref{fig:ML_mdot}, HFS are consistent with accretion rates of $\dot{m}\gtrsim 200-300$~M$_\odot$~pc$^{-1}$~Myr$^{-1}$, that is much higher than those expected for most ISM filaments (Sect.~\ref{sec:accretion_filaments}).

Previous observational works have independently investigated the total accretion rate in HFS using 
mass flows along filaments feeding the HFS \citep[e.g.,][]{Kirk2013, Peretto2013,Hacar2017a,Williams2018} and/or infall motions along the line of sight associated to the gas around these objects \citep[e.g.,][]{Walsh2006,Schneider2011,Yang2023_HFS}.
We have compiled some of these studies in Table~\ref{table:HFS_data} (see references therein)\footnote{We note that most accretion measurements are affected by projection effects and are therefore subject of relatively large uncertainties \citep[see][for a discussion]{Peretto2014}.}.
Accretion rates in HFS are estimated to be in the range between $\dot{M}_{HFS}\sim 20-4500$~M$_\odot$~Myr$^{-1}$ with a mean value of $\dot{M}_{HFS}\sim 1000$~M$_\odot$~Myr$^{-1}$. For comparison, the typical accretion rates onto dense cores at scales of $\sim$~0.1~pc is estimated between $\dot{M}\sim 10$~M$_\odot$~Myr$^{-1}$ \citep{Lee2001_infall} and $\dot{M}\sim 100$~M$_\odot$~Myr$^{-1}$ \citep{Wells2024}.

We find a direct correlation between the observed HFS masses $M_{HFS}$ and their corresponding accretion rates $\dot{M}_{HFS}$. We illustrate this correlation in Fig.~\ref{fig:HFS_mdot} (upper panel) where a fit to these data retrieves an almost linear proportionality between these quantities following $\dot{M}_{HFS}\propto M_{HFS}^{1.11}$ (red dashed line). The above correlation is steeper than the one found for massive clumps \citep{traficante2023,Zhou2024} and cores \citep{Wells2024}.

We have converted the above $\dot{M}_{HFS}$ values into equivalent accretion rates per unit length $\dot{m}_{HFS}$ assuming $\dot{m}_{HFS}=\dot{M}_{HFS}/L_{HFS}$.
In Fig.~\ref{fig:ML_mdot}, we color-coded the above estimated $\dot{m}_{HFS}$ rates similar to our model predictions (see color bar).
As expected, observations report accretion rates between $\dot{m}_{HFS}\gtrsim 50-2000$~M$_\odot$~pc$^{-1}$~Myr$^{-1}$ in HFS of $M>100$~M$_\odot$.

We further illustrate the relative good agreement between these estimates in Fig.~\ref{fig:HFS_mdot} (lower panel) where we display the individual observed mass accretion rates $\dot{m}_{HFS}$ against the corresponding predictions for $\dot{m}_{pred}$ from Eq.~\ref{eq:mdot_pred} 
given the $M_{HFS}$ and $L_{HFS}$ values in Table~\ref{table:HFS_data}.
We note that the predictions derived from Eq.~\ref{eq:mdot_pred} only assume that the accretion timescale is equal to the fragmentation timescale (Eq.~\ref{eq:tau_acc_vs_frag}) as working hypothesis of our toy model (Sect.~\ref{sec:toymodel}). 
We observe a positive correlation between $\dot{m}_{HFS}$ and $\dot{m}_{pred}$ (dashed grey line; {\bf $\dot{m}_{HFS}\propto \dot{m}_{pred}^{1.13}$}), although our model seems to slightly overpredict the actual accretion rates by a factor of $\sim$3.
Similar to Fig.~\ref{fig:HFS_mdot} (upper panel), a comparison between the observed $m_{HFS}$ and $\dot{m}_{HFS}$ (not shown) also retrieves a positive and tight correlation between these observables close to our expectations where $\dot{m}_{HFS}\propto m_{HFS}^{1.15}$, although again with some scatter.
The obvious limitations of our toy model together with observational biases such as projection effects and large uncertainties on the mass estimates can hamper these comparisons. Given its simplicity and basic assumptions, however, this close correspondence between these observed and predicted accretion rates reinforces the validity of these estimates, at least in a first order approximation.


\section{Inducing gravitational collapse}\label{sec:inducingcollapse}


\begin{figure}[t]
\centering
\includegraphics[width=1.\linewidth]{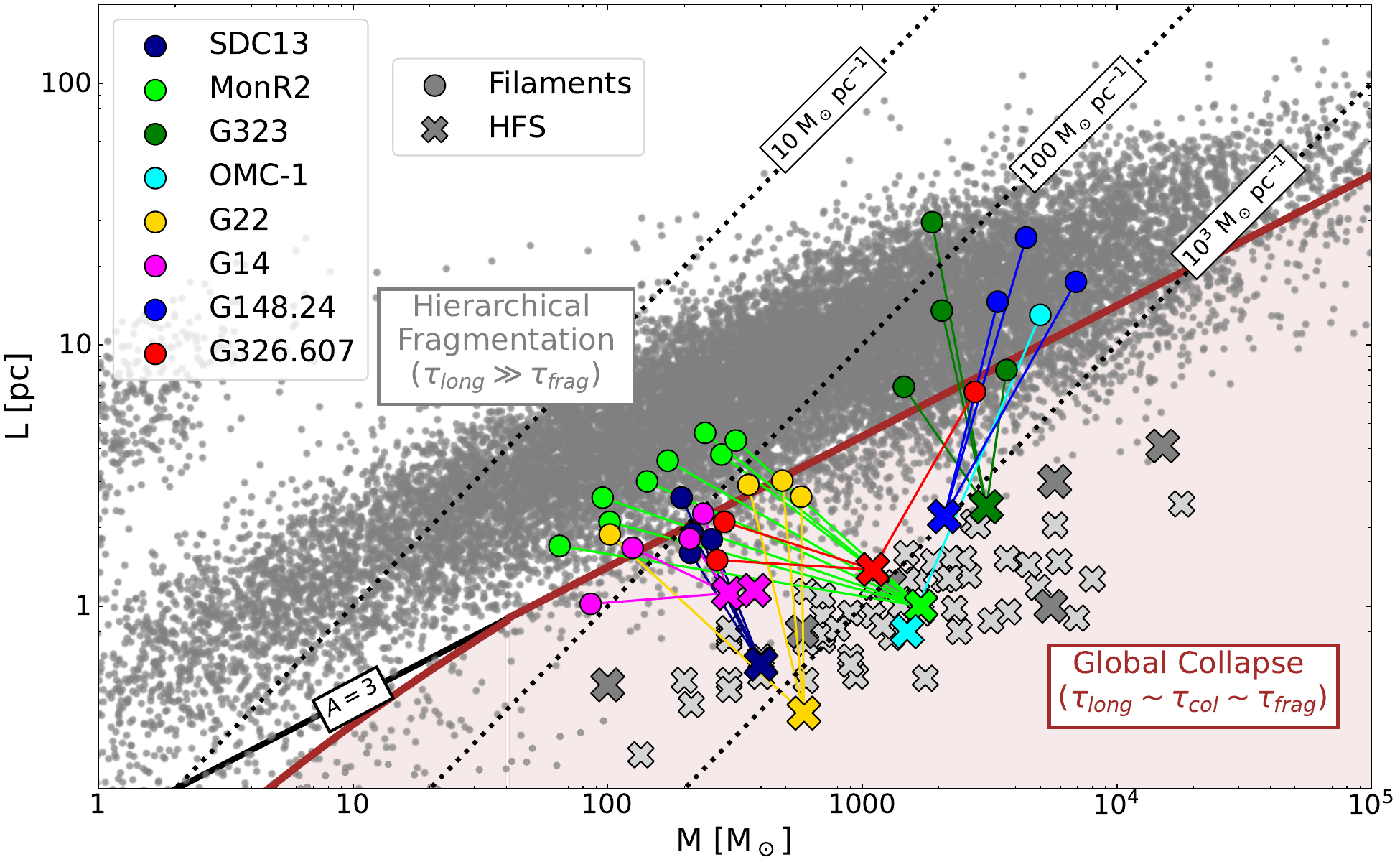}
      \caption{Mass-Length plot comparing those HFS (crosses) with their pc-scale filaments (circles). Individual regions, where filaments and HFS are connected by segments, are color coded in the plot (see legend; see references in Table~\ref{table:HFS_data}).
      Lines and labels are similar to Fig.\ref{fig:ML_mdot}.
      We highlight different diagonal lines defining individual $m=M/L$ values (=[10,100,100]~M$_\odot$~pc$^{-1}$; dotted black lines).
      }
\label{fig:hub_vs_fil}
\end{figure}

\subsection{A geometric phase-transition}\label{sec:phasetrans}

Mergers and collisions between filaments 
\citep{Duarte2011,Nakamura2014,JimenezSerra2014,Beltran2022,Kumar2022}, and more generally converging flows \citep{Heitsch2008,Schneider2010,GalvanMadrid2010} and cloud-cloud collisions \citep{Higuchi2010,Fukui2014,Wang2022_HFS,Zhou2023}, have been suggested as the potential origin of parsec-size HFS leading to the formation of clusters in the Solar neighbourhood \citep{Nakamura2012,Duarte2010}, massive IRDCs \citep{Schneider2010,GalvanMadrid2013,Panja2023}, as well as superclusters in both the Milky Way \citep{Fukui2016} and the Magellanic Clouds \citep{Fukui2015,Saigo2017}, to cite few examples.
Different kinematic \citep{Duarte2011,Torii2017,Haworth2015} and chemical \citep{JimenezSerra2010,Sanhueza2013} features have been proposed as observational signatures of these collisions.
Kinematic studies reveal clear signatures of global inward motions as signature of the gravitational collapse of these HFS after formation \citep{Myers2009a,Peretto2013,Hacar2017a,Williams2018,He2023,Zhou2024,Zhang2024_HFS}.
As expected for gravitationally dominated systems, magnetic fields appear to be sub-dominant in HFS once their collapse has started \citep{Pattle2017,Wang2020c,Arzoumanian2021}.
Complementing these works, our toy model provides a simple and testable analytic framework aiming to capture the main physical properties of these HFS.

Compared to filaments, and according to Eq.~\ref{eq:tau_long_vs_frag}, $\tau_{long}$ becomes comparable to $\tau_{frag}$ in more prolate geometries \citep[$A< 3$; ][]{Pon2012,Toala2012} and therefore should be considered for their evolution ($\tau_{long}\lesssim \tau_{frag}$). 
For regions where also $m>m_{vir}$ (i.e. supervirial) the radial collapse timescale ($\tau_{col}\sim \tau_{ff}$) becomes comparable to the fragmentation timescale \citep[$\tau_{col}\sim \tau_{frag}$; e.g.,][]{Inutsuka1997}. 
Filamentary clouds located towards the right, lower corner of Fig.\ref{fig:ML_mdot} which simultaneously satisfy both conditions, that is are spheroidal with $A\lesssim 3$ and are supervirial with $m>m_{vir}$, are therefore prone for collapse.
This transition from a filamentary ($A>3$) to a spheroidal ($A\lesssim 3$) geometry can therefore induce the global (longitudinal + radial) collapse of a object since in these regions all timescales are comparably fast, i.e. $\tau_{long}\sim\tau_{frag}\sim \tau_{col}$. We have indicated the parameter space for this (gravitational) global collapse in Fig.\ref{fig:ML_mdot} (shaded brown area) defined as the area where both spheroidal and supervirial conditions are satisfied (see also Fig.~\ref{fig:ML_AR}).

A geometric phase-transition between a filamentary and a spheroidal ($A=3$) configuration could explain the origin of HFS. As shown in Fig.~\ref{fig:ML_mdot}, HFS are identified as supervirial structures ($m>m_{vir}$) with sphere-like aspect ratios ($A<3$) showing large mass accretion rates ($\dot{m}\gtrsim 300$~M$_\odot$~pc$^{-1}$~Myr$^{-1}$,  Sect.~\ref{sec:accretion}). Unlike filaments, these objects satisfy the conditions for global collapse on scales of 0.5-2 pc in timescales of $\sim$1~Myr (see above). We notice that, although not required by our analytic calculations, the transition into collapse at $A\leq 3$ occurs at an average surface density of $\Sigma \geq 8\times 10^{21}$~cm$^{-2}$ (or A$_V~\sim$~8~mag) coincident with the threshold for star formation observed in nearby clouds \citep{Lada2010}.

The idea of induced gravitational collapse after merging (see models and references above) is reinforced by the comparison between the properties of the HFS and their individual filaments. In Fig.~\ref{fig:hub_vs_fil} we display those regions in our sample for which mass and length measurements of their connecting, pc-scale filaments are available (color coded, see legend; see also references in Table~\ref{table:HFS_data}). As discussed in Sect.~\ref{sec:AR_col}, the predicted line for $A=3$ separates these (sub-)virial filamentary structures from their supervirial HFS also in individual targets.
In all cases the local line mass (see dotted black lines) of the HFS ($m_{HFS}$) is also $>3$ times larger than the corresponding line mass of any of their individual filaments $m_{f}$, even in extreme cases such as DR21 \citep{Schneider2011,Hennemann2012} or Mon-R2 \citep{Trevino-Morales2019,Rayner2017}. As discussed by \citet{Hacar2022}, the segmentation of a filament would preserve its line mass leading to changes along constant $m$ values and consequently $m_{HFS}\sim m_{f,i}$. On the contrary, correlations such as $m_{HFS} > m_{f,i}$ indicate that HFS locally ($\lesssim$~1~pc) gathered additional mass via merging of filaments (e.g. $m_{HFS,0}\sim\sum_i m_{f,i}$) which in turns triggers large accretion rates (Sect.~\ref{sec:accretion}) due to the induced gravitational collapse of these supervirial regions.

The observed HFS mass $M_{HFS}$ is expected to combine the contributions of the initial HFS mass $M_{HFS,0}$ and the mass accreted over time $\tau$ such as $M_{HFS}(\tau) = M_{HFS,0}+\dot{M} \cdot \tau$ or, similarly, $ m_{HFS}(\tau) = m_{HFS,0}+\dot{m}_{HFS} \cdot \tau$, if written in terms of line mass.
In a simplified configuration in which a newly formed HFS condenses all the mass of the merger, so $m_{HFS,0}\sim \sum_i m_{f,i}$, the line mass of a HFS over time could be described as
\begin{equation}\label{eq:mfinal}
    m_{HFS}(\tau) \simeq \sum_i m_{f,i} +\dot{m}_{HFS} \cdot \tau.
\end{equation}
The final value of $m_{HFS}$ can be further amplified by the collapse of these regions reducing L.
The results shown in Fig.~\ref{fig:hub_vs_fil}, where several targets show $m_{HFS} > \sum_i m_{f,i}$, suggest that the currently observed mass accretion rates after merging $\dot{m}_{HFS}$ (Sect.~\ref{sec:accretion}) could significantly contribute to the mass of these regions within the typical evolutionary times for HFS of $\tau~\sim $~1-2~Myr \citep{Myers2009a}.

\subsection{Filamentary vs spherical accretion}

Increasing observational evidence demonstrates the presence of significant filamentary accretion flows onto HFS \citep[e.g.,][]{Kirk2013,Chen2019,Baug2018,Dewangan2020,Ma2023,Panja2023,Rawat2024}.
These findings led to the proposal of formation scenarios for high-mass stars involving multiple-scale processes in HFS such as the clump-fed \citep{Tige2017,Motte2018,Peretto2020} and the inertial-flow  models \citep{Padoan2020}.

Following a simplified merging scenario (Sect.~\ref{sec:phasetrans}), our description assumes that most of the mass in a HFS can be originally associated to filamentary structures at larger scales. Different CO surveys included in our HFS sample such as G326 \citep{He2023}, G323.46 \citep{Ma2023}, G6.55 \citep{Sen2024}, or G148.24 \citep{Rawat2024}, as well as other studies targeting HFS \citep{Anderson2021,Peretto2023}, 
appear to support this working hypothesis. Yet, the actual gas organization inside and outside HFS is likely more complex. The gas in molecular clouds is known to be multi-fractal and combines diffuse (gaussian) and filamentary (non-gaussian) contributions at all scales \citep{Robitaille2020}. HFS may thus accrete material from both filaments and diffuse gas simultaneously. 

The distinction between these collimated (filamentary) and spherical (diffuse gas) accretion modes in HFS is customary in other astrophysical problems such as the study anisotropic accretion onto galaxies via cold streamers \citep[e.g.,][]{Dekel2009_coldstreams}.
The relative importance of these two accretion modes can be evaluated using geometrical considerations.
The total mass accretion $\dot{M}$ can be parameterized from the gas mass flux over the solid angle $\Omega$ produced by gas at density $n$ with an accretion velocity $v_{acc}$, such as $\dot{M}=n\cdot \Omega \cdot v_{acc}$ \citep[e.g.,][]{Kirk2013,Peretto2013,Hu2021}.
At a radius $r$ from the centre of a HFS, the total filamentary accretion funnelled by $N_{f}$ filaments of radius $R_{f}$, cross section $\Theta_f=\pi R_{f}^2$, and density $n_f$, over a solid angle $\Omega_{f}=\frac{\Theta_f}{r^2}=\pi \left( \frac{R_f}{r}\right)^2$ is then 
\begin{equation}
    \dot{M}_{f}=n_{f}\cdot N_{f}\ \Omega_{f}\cdot v_{acc}.
\end{equation}
On the other hand, the maximum spherical accretion produced by the background gas of density $n_{bg}$ outside filaments can be estimated as 
\begin{equation}
\dot{M}_{bg}=n_{bg}\cdot (4\pi-N_{f}\ \Omega_{f})\cdot v_{acc}.
\end{equation}
Assuming that both gas components experience the same accretion velocity $v_{acc}$ (e.g., free-fall), the relative contribution of these filamentary and spherical accretions to the evolution of the HFS can be evaluated as
\begin{equation}\label{eq:difacc}
    \frac{\dot{M}_{bg}}{\dot{M}_{f}}=\frac{n_{bg}\cdot (4\pi-N_{f}\ \Omega_{f})}{n_{f}\cdot N_{f}\ \Omega_{f}}
\end{equation}
To be dynamically relevant (i.e. $\dot{M}_{bg}\sim \dot{M}_{f}$), the density of the diffuse gas around the HFS must be
\begin{equation}\label{eq:difacc2}
    n_{bg}\sim\frac{n_{f}}{\left(\frac{4\pi}{N_{f}\ \Omega_{f}} -1\right)}=\frac{n_{f}}{\left(\frac{4\ r^2}{N_{f}\ R_{f}^2} -1\right)}.
\end{equation}
Using Eq.~\ref{eq:difacc2}, it is easy to demonstrate that at the short radii $r\lesssim 2 R_{f}$ ($\sim R_{HFS}$) expected at the intersection of $N_{f}=2$ filaments, the density of the diffuse gas needs to be comparable to the gas density in filaments ($n_{bg}\gtrsim n_{f} / 7$) to influence the mass growth of HFS. 
This background density should be even higher if more filaments $N_{f}>2$ contribute to the filamentary inflow. On the other hand, diffuse accretion may become more relevant at larger radii outside the central HFS where the minimum density contrast required by Eq.~\ref{eq:difacc2} decreases rapidly as $\frac{n_{bg}}{n_{f}}\propto \frac{1}{r^2}$ at $r\gg R_f$.

As discussed in Sect.~\ref{sec:accretion_filaments}, diffuse gas accretion determines the evolution and mass load of individual filaments. On the contrary, the accretion of diffuse gas onto HFS may play a secondary role with respect to the mass accreted throughout filaments ($\dot{M}_{f}>\dot{M}_{bg}$) unless embedded in a high density environment.
Given the high column density contrast seen in HFS \citep{Myers2009a,Kumar2020}, mass accretion in HFS is then expected to be dominated by filamentary flows, in agreement to simulations \citep{Gomez2014,VazquezSemadenietal2019}.


\subsection{A dichotomy in the M-L phase-space}\label{sec:evolution}

\begin{figure*}[th!]
\centering
\includegraphics[width=1.\linewidth]{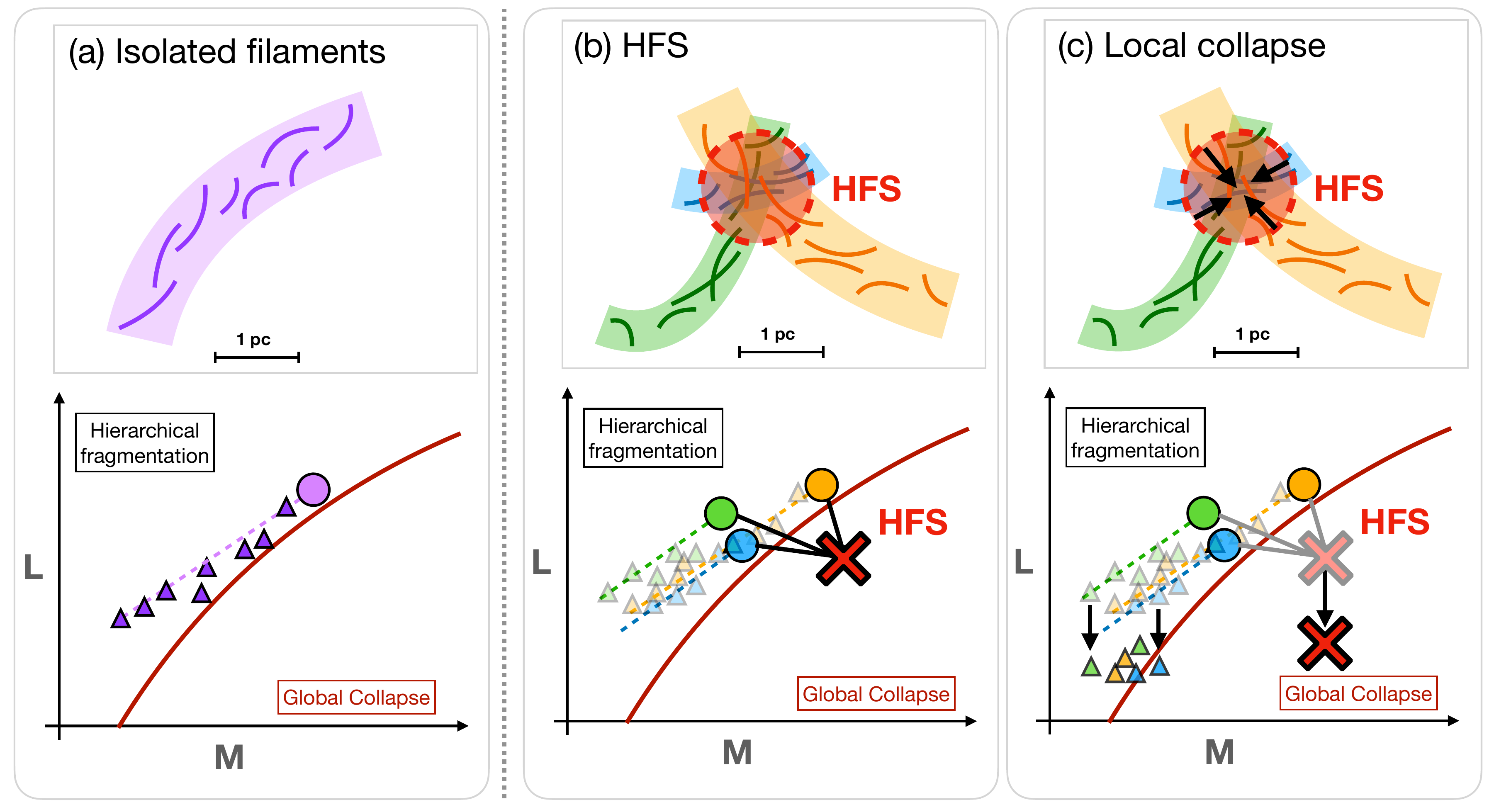}
      \caption{Illustrative cartoon describing the configuration (top panels) and location of filaments and HFS in the M-L phase-space (bottom panels). 
      From left to right: 
      {\bf(a)} hierarchical fragmentation of isolated filaments (circles) into sub-filaments (triangles), 
      {\bf(b)} formation of a HFS system at the intersection of multiple filaments, and
      {\bf(c)} local gravitational collapse of the resulting parsec-scale clump. Black arrows indicate how this collapse may lead into the shrink in size of both the central clump and sub-filaments in the HFS moving these objects downwards in the M-L phase-space. Similar to Fig.~\ref{fig:HFS_mdot}, a brown line in these plots indicates the transition between hierarchical fragmentation (above the line) and global collapse (beneath the line).
      }
\label{fig:HFS_cartoon}
\end{figure*}

In addition to previous formation models (see Sect.~\ref{sec:phasetrans}),
we aim to explain how HFS may evolve outside the main sequence of filaments seen in the M-L plane generated by their turbulent fragmentation (Sect.~\ref{sec:accretion_filaments}) and populate a different regime of this phase-space (see Fig.~\ref{fig:ML_phasespace}).
To facilitate our description, we summarize these differences in three characteristic configurations shown by corresponding cartoons in Fig.~\ref{fig:HFS_cartoon} motivated by our previous observational results and models (Figs.~\ref{fig:ML_mdot}-\ref{fig:hub_vs_fil}).
These diagrams are made for illustrative purposes only, acknowledging the strong (over-)simplifications of these cartoons. 
Detailed simulations are needed to explore more realistic, self-consistent evolutionary tracks for different filament and HFS configurations \citep[e.g.,][]{Feng2024}.

Multiple mechanisms generate parsec-size, elongated ($A\gg 3$) structures as part of the intrinsic filamentary nature of the ISM \citep[see][and references therein]{Hacar2022}. Usually showing a low column density ($\Sigma_f$) and subvirial state ($m_f<m_{vir}$), these parsec-size filaments (circles) typically fragment creating a hierarchical sub-structure of smaller filaments (triangles) that move diagonally towards the lower left corner of this M-L diagram, following a relation such as $L\propto M^{0.5}$ (Eq.\ref{eq:ML}) (Fig.~\ref{fig:HFS_cartoon}, panel a;  see also Sect.~\ref{sec:accretion_filaments} and Fig.~\ref{fig:ML_cartoon}). 
Given their relatively low accretion rates $\dot{m}\sim 10-100$~M$_\odot$~pc$^{-1}$~Myr$^{-1}$, isolated filaments are therefore unable to reach the conditions for the formation of high-mass stars. 

On the other hand, HFS are created at the junction of several of these parsec-size filaments (Fig.~\ref{fig:HFS_cartoon}, panel b). With typical cross sections of $\sim$~1~pc, these filament junctions generate prolate regions ($A\leq 3$), which rapidly increase both the total (M$_{HFS}$) and line ($m_{HFS}>m_f$) masses as well as the clump surface density ($\Sigma_{HFS}>\Sigma_f$), shifting the HFS towards the lower-right direction in the M-L diagram with respect to their initial filaments ($m_{HFS}>m_f$) as shown in Fig.~\ref{fig:hub_vs_fil}. 

If the merging of filaments is efficient enough, the large accretion rates $\dot{m}\gtrsim 300$~M$_\odot$~pc$^{-1}$~Myr$^{-1}$ in these junctions (Sect.~\ref{sec:accretion}) can make them locally supervirial ($m_{HFS}>m_{vir}$) and the resulting HFS would then collapse by its own gravity ($\tau_{frag}\sim \tau_{long}$; Sect.~\ref{sec:inducingcollapse}). Shrinking in size and gaining mass (while also fragmenting inside, triangles to the left), this runaway process may continue moving the HFS downwards in the M-L phase space away from the main distribution of filaments (Fig.~\ref{fig:HFS_cartoon}, panel c). This global collapse may explain the apparent gap between HFS and filaments in the M-L phase-space.
Other gas structures inside these HFS (e.g., cores and sub-filaments) would likely experience additional compression induced by the parsec-scale collapse (moving down in this M-L diagram).  
Unlike any filament fragmentation process (where $m_f< m_{vir}$), the collapse induced during the transition from elongated to prolate geometries marks a physical change (phase-transition) on the evolution of these HFS and the only way to agglomerate enough mass to form large stellar clusters ($M_\star>100$~M$_\odot$), and eventually high-mass stars, in timescales of $\lesssim$~1~Myr \citep{Motte2018}.

Panels b and c in Fig.~\ref{fig:HFS_cartoon} resemble the early stages I plus II proposed by \citet{Kumar2020} for the evolution of HFS. In addition to this previous qualitative description, our model provides the physical foundations for some of the early steps on the evolution of HFS as well as direct quantitative predictions for future observations (i.e., $M$, $L$, $A$, $\Sigma$, and $\dot{m}$). Our analytic description (Sect.~\ref{sec:toymodel}) also suggests a dichotomy between the dominant physical mechanisms governing the evolution isolated filaments and HFS.
Individual, parsec-size filaments primarily evolve in a top-down manner  (Sect.~\ref{sec:accretion_filaments}) dominated by an internal turbulent fragmentation \citep[{\it fray-and-fragment model};][]{Tafalla2015} populating the main diagonal of the M-L phase-space (Fig.~\ref{fig:HFS_cartoon}, panel a). On the other hand, HFS might only emerge from a fast, bottom-up process triggered by the agglomeration of filaments (Sect.~\ref{sec:inducingcollapse}) and the gravitational collapse of these regions on parsec scales \citep[{\it gather-and-fray model};][]{Smith2016} (Fig.~\ref{fig:HFS_cartoon}, panels b and c).

In models of global hierarchical collapse \citep[e.g.,][]{VazquezSemadenietal2019,Camacho2023}, HFS appear as the manifestation of a fast local collapse within a the much slower large-scale contraction of gravitationally bound clouds. 
A multi-scale organization of HFS including gas flows from cloud to core scales has been proposed as the origin of these systems in different high-mass star-forming regions \citep{Zhou2023,Zhou2024_galaxy}.
Instead, the observed separation in the M-L phase-space appears to not only separate the HFS from their parental filaments but suggests that only the former may satisfy the supercritical conditions ($m>m_{vir}$) for local collapse. Restricted to a volume comparable to the initial merger cross-section, the gravitational collapse in HFS is likely to be limited to a few parsecs within the lifetime of this structures. Our analysis supports recent results indicating that the gas clumps leading to the formation of clusters, analogous to our HFS (Sect.~\ref{sec:HFSclusters}), may be dynamically decoupled from their molecular clouds at scales of few parsecs \citep{Peretto2023}. The transition to global collapse corresponds with a change on the virial properties ($\alpha_{vir}\lesssim 1$) of these clumps observed at $\Sigma\sim 300$~M$_\odot$~cm$^{-2}$ (or $\Sigma\sim 1.4\times 10^{22}$~cm$^{-2}$) \citep{Kauffmann2013,Traficante2020,Peretto2023} in close agreement with our analytic predictions for HFS with $A=1.5$ and $\Sigma\sim 1.6\times 10^{22}$~cm$^{-2}$ (see Eq.~\ref{eq:a_Sigma} and Fig.~\ref{fig:ML_AR}).
Our results favor the ideas introduced by the inertial-flow model for star formation \citep{Padoan2020}, although applied to parsec-size, cluster-forming HFS.

\section{A scale-free process?}\label{sec:scalefree}

Given the fractal, scale-free nature of the filamentary ISM \citep{Falgarone1991,Elmegreen1996,Hacar2022}, it would be interesting to explore whether the same transition between filamentary and spherical geometries could also explain the formation of distinct collapsing regions at different scales. High-resolution observations report the existence of hub-like structures created by the interaction of small filaments (aka fibers) at sub-parsec scales in regions such as SDC335-MM1 \citep{Peretto2013,Xu2023}, NGC~1333 IRAS~3 \citep{Hacar2017b}, G035.39 \citep{Henshaw2014}, OMC-1 Ridge and OMC-2 FIR-4 \citep{Hacar2018}, and OMC-3 MM7 in Orion \citep{Ren2021} currently forming the most massive protostars in these clouds. Translated to smaller masses of M~$\lesssim$~100~M$_\odot$, the same physical mechanism operating in the parsec-size HFS discussed above could also explain the origin of overdensities at sub-parsec scales if these (mini-)hubs would become gravitationally unstable (by entering the shaded area in Fig.\ref{fig:ML_mdot}).   

Particularly relevant for this discussion is the origin of high-mass cores ($M\sim$10-100~M$_\odot$). 
Theoretical considerations derive mass accretion rates onto massive stars of $\dot{M}_\star= 1-6\times 10^{-4}$~M$_\odot$~yr$^{-1}$ \citep{McKeeTan2002}, consistent with the infall values derived for their parental star-forming clumps at scales of $\sim$~0.1~pc \citep{Fuller2005}, that is, $\dot{m}\sim 1-6\times 10^3$~M$_\odot$~pc$^{-1}$~Myr$^{-1}$. As shown in Fig.\ref{fig:ML_mdot}, the line $\dot{m}= 10^3$~M$_\odot$~pc$^{-1}$~Myr$^{-1}$ almost coincides with the relation defining the mass threshold for high-mass star formation with $M_{HM}=870\ M_\odot\cdot (R_{eff}/pc)^{1.33}$ \citep[dashed purple line;][]{KauffmannPillai2010} and similar revised estimates \citep[$M_{HM}=1282\ M_\odot\cdot (R_{eff}/pc)^{1.42}$;][]{Baldeschi2017}. 
In analogy to the large HFS, the merging of filaments at sub-parsec scales inducing junctions (or forks) with masses below this line could trigger the formation of such massive cores, a possibility currently explored in simulations \citep{Smith2011,Clarke2017,Hoemann2021,Kashiwagi2023,Hoemann2024}. 
While locally collapsing, these massive cores could still fragment and form a binary (multiple) star system at smaller scales.

Continuing towards lower masses, the process could also be extended to dense (solar-like) cores with typical sizes L$\sim$~0.1~pc and masses M~$\sim$~3~M$_\odot$ \citep{Myers1983}. Formed out of the quasi-static fragmentation of some of sub-pc filaments at the end of the turbulent cascade \citep{Hacar2011}, the change in geometry from filamentary into a prolate structure \citep[with $A=1.5$, see dashed orange line in Fig.~\ref{fig:ML_AR};][]{Myers1991_prolate} could trigger local collapse in dense cores if $\tau_{long}\sim \tau_{col}$ \citep[e.g.,][]{Heigl2016}. A comparison between our models and the average mass and length properties \citep[M~$\sim$~4.1~M$_\odot$ and L$\sim$~0.12~pc;][]{Myers1983} as well as accretion rates in dense cores \citep[$\dot{M}> 10$~M$_\odot$~Myr$^{-1}$ or $\dot{m}> 100$~M$_\odot$~pc$^{-1}$~Myr$^{-1}$,][]{Lee2001_infall} appear to support this hypothesis. 

The lower mass-end of the area defining global collapse is shaped by the sonic (thermal) limit of Eq.\ref{eq:mvir} when $m_{vir}\sim \frac{2c_s}{G}$. Low-mass ($<2$ M$_\odot$) spheroidal ($A<3$) structures could therefore exist without necessarily collapsing (see lower left corner of the M-L phase-space). This could explain the existence of small, pressure-confined but still starless cores reported in different star-forming regions \citep{Lada2008,Kirk2017}. In this low-mass regime, comparatively lower L values (i.e.
increasingly higher densities following Eq.~\ref{eq:nL}) are needed to trigger collapse as expected by the Jeans criterion ($M_J\propto n^{-1/2}$). 
Further observational and theoretical studies are needed to explore the tantalizing ideas presented in this section.

\section{Conclusions}
In this Paper V we introduce an analytic formulation to explore the origin and early evolution of hub-filament systems (HFS)  within of the filamentary structure of the ISM. The results of our paper can be summarized as follow:
\begin{enumerate}
    \item  We investigated the properties of HFS with masses $M_{HFS}\sim 100-2\times10^4$~M$_\odot$ and sizes up to $L_{HFS}\sim 5$~pc as likely precursors of typical Milky Way clusters with up to few thousands solar masses in stars ( Sect.~\ref{sec:toymodel}).
    
    \item The enhanced mass in HFS set these objects apart from the fundamental M-L scaling relation ($L\propto M^{0.5}$) governing the filamentary structure of the ISM at similar scales.
    
    \item  Using a simple analytic formulation, we created a toy model to explore the evolution of different filamentary structures in the M-L phase-space. Despite its simplicity, our toy model provides direct predictions of the filament aspect ratio ($A$), average gas column density ($\Sigma$), relevant timescales (i.e., $\tau_{col}$, $\tau_{long}$, $\tau_{frag}$, $\tau_{acc}$), and accretion rates ($\dot{m}$) in close agreement with observations.
    
    \item  Our model predicts an inverse proportionality between the aspect ratio $A$ and total gas surface density $\Sigma$ of filaments at different scales. Once fixed, the value of $A$ (or $\Sigma$) determines the normalization value $a$ for their expected $L= a\cdot M^{0.5}$ correlation. 
    
    \item Most of the molecular filaments in the ISM are bracketed between the model predictions for $A\leq 30$ (or $\Sigma= 8\times 10^{20}$~cm$^{-2}$) and $A\geq 3$ (or $\Sigma=8\times 10^{21}$~cm$^{-2}$).
    On the other hand, HFS occupy the expected parameter space for $A\lesssim 1.5$ (or $\Sigma\gtrsim 1.6\times 10^{22}$~cm$^{-2}$). Our results suggests that the observed differences between filaments and HFS may arise from the spheroidal geometry of these later objects.
    
    \item Our model predicts a dichotomy between the dominant physical mechanisms governing the evolution of isolated filaments and HFS (Sect.~\ref{sec:timescales}).
    Filamentary structures with large aspect ratios ($A>3$) fragment in timescales much shorter than their longitudinal contraction ($\tau_{frag}<\tau_{long}$). In contrast, the combination of high line masses ($m>m_{vir}$) and spheroidal shapes ($A<3$) seen in HFS makes these structures prone to global (radially + longitudinally) collapse ($\tau_{frag}\sim\tau_{long}\sim \tau_{col}$) on timescales of $\sim$~1~Myr (Sect.~\ref{sec:timescales}).
    
    \item Our analytic calculations predict the fast evolution of HFS to be favored by the high accretion rates ($\dot{m}\gtrsim 300$~M$_\odot$~pc$^{-1}$Myr$^{-1}$) originated at the junction of multiple parsec-scale filaments.
    Despite its simplicity, our model predictions reproduce the observed variations of {$\dot{M}$ across more than an order of magnitude in $M$ in HFS.}
    
    \item The change from a filamentary into spheroidal gas organization may induce geometrical phase-transition triggering the (global) gravitational collapse of the HFS at parsec scales (Sect.~\ref{sec:inducingcollapse}). This transition seems to occur at column densities of $\Sigma\sim8\times 10^{21}$~cm$^{-2}$, coincident with the previously proposed threshold for star formation.
    
    \item This gravitational collapse would make these HFS to shrink in size detaching these objects from most ISM filaments in the M-L phase-space and setting the initial conditions for the formation of high-mass stars and clusters.
    
    \item We speculate that a similar process might also operate at sub-parsec scales, explaining the formation of similar mini-hubs and cores connected to the formation of individual stars (Sect.~\ref{sec:scalefree}).
\end{enumerate}

\begin{acknowledgements}
      The authors thank D. Arzoumanian for kindly sharing their data for this analysis.
      A.H., F.B., and A.S. received funding from the European Research Council (ERC) under the European Union’s Horizon 2020 research and innovation programme (Grant agreement No. 851435).
      DS acknowledges support of the Bonn-Cologne Graduate School, which is funded through the German Excellence Initiative as well as funding by the Deutsche Forschungsgemeinschaft (DFG) via the Collaborative Research Center SFB 1601 ``Habitats of massive stars across cosmic time'' (subprojects B4 and B6).
      Furthermore, the project is receiving funding from the programme
      ``Profilbildung 2020'', an initiative of the Ministry of Culture and Science of the State of Northrhine Westphalia.
      E.S. acknowledge financial support from  European Research Council via the Horizon 2020 Framework Programme ERC Synergy ``ECOGAL'' Project GA-855130.
      This work made use of Astropy:\footnote{http://www.astropy.org} a community-developed core Python package and an ecosystem of tools and resources for astronomy \citep{astropy:2013, astropy:2018, astropy:2022}. This work made intensive use of the python packages {\it Matplotlib} \citep{matplotlib:2007} and {\it Numpy} \citep{Numpy:2020}.
\end{acknowledgements}
\bibliographystyle{aa}
\bibliography{filaments.bib}

\begin{appendix}
\section{Literature results}\label{sec:appendixA}
We have compiled the mass (M), length (L), and mass accretion rate ($\dot{m}$) for those parsec-scale HFS available in the literature in Table~\ref{table:HFS_data}.

\begin{table*}[ht!]
\caption{Hub-Filament Systems (HFS) at parsec scales: Literature results ordered by mass.}             
\label{table:HFS_data}      
\centering          
\begin{tabular}{l c c c c l }     
\hline\hline       
HFS & $M_{HFS}$ & $L_{HFS}$ $^{(1)}$ & $\dot{M}_{HFS}$ $^{(2)}$ & $\dot{m}_{HFS}$ $^{(3)}$   & References \\ 
  & [M$_{\odot}$] & [pc] & [M$_{\odot}$~Myr$^{-1}$] & [M$_{\odot}$~pc$^{-1}$~Myr$^{-1}$]  &  \\ 
  \hline
  \multicolumn{6}{c}{targets with $\dot{m}$ estimates}\\
\hline            
  \object{DR21} & 15120 & 4.1 & 4150 & 1100 & \citet{Schneider2011,Hu2021} \\
  SDC335.579-0.272 & 5500 & $\sim$1 & 2500 & $\sim$2500 & \citet{Peretto2013} \\
  G6.55-0.1 & 4520 & 7.6 & 1778 & 234 & \citet{Sen2024}	\\
  G323.46-0.08 & 3072 & 2.4 & 1280 & 506 & \citet{Ma2023} \\
  G148.24+00.41 & 2100 & 2.2 & 675 & 306 & \citet{Rawat2024} \\
  \object{Mon-R2} & $>$1700$^{(4)}$ & $\sim$1 & 500 & $\sim$500 & \citet{Trevino-Morales2019} \\
  \object{OMC-1} & $>$1500$^{(4)}$ & 0.8 & 385 & 481 & \citet{Hacar2017a} \\
  G310.142+0.758 & 1280 & 0.6 & 2400 & 2166 & \citet{Yang2023_HFS} \\
  G326.607+00.799 & 1099 & 1.38 & 170 & 123 & \citet{He2023} \\	
  G22-C1 & 590 & 0.39 & 360 & 923 & \citet{Yuan2018} \\
  \object{NGC~1333} & 580 & 0.8 & 100 & $\sim$125 & \citet{Hacar2017b,Walsh2006} \\
  SDC13 & $\sim$400 & $\sim$1 & 15 & 25 & \citet{Peretto2014} \\ 
  G14.225-0.506 South & 377 & 1.15 & 130 &  89 & \citet{Busquet2013,Chen2019} \\
  G14.225-0.506 North & 297 & 1.12 & 100 & 116 & \citet{Busquet2013,Chen2019} \\
  \object{Aquila-Rift} & $>$100 & 0.5 & 30 & 60 & \citet{Kirk2013} \\
  \hline
  \multicolumn{6}{c}{surveys without $\dot{m}$ estimates}\\
  \hline
  12$\times$ HFS & 1353-18$\times$10$^4$ & 0.9-2.46 & --- & --- & \citet{Tokuda2023} \\
  39$\times$ HFS & 200-4900 & 0.48-1.52 & --- & --- & \citet{Morii2023} \\
  5$\times$ HFS & 135-3739 & 0.27-0.95 & --- & --- & \citet{Anderson2021} \\

\hline                  
\end{tabular}
\tablefoot{(1) When the HFS radius $R$ is provided we estimate the corresponding length as $L_{HFS}=2R$.
(2) Most accretion measurements are affected by projection
effects and are therefore subject of relatively large uncertainties \citep[see][, for a discussion]{Peretto2014}.
(3) Estimated as $\dot{m}_{HFS} = \dot{M}_{HFS}/L_{HFS}$.
(4) Lower mass limit due to the large contribution of stellar objects.
}
\end{table*}

\section{Alternative derivation}\label{ap:alternative}

Equations \ref{eq:surfden_AR}-\ref{eq:a_Sigma} can be also derived from the expected correlation between mass and density with in gas surface density in connection with different scaling relations.
In particular, Eq.~\ref{eq:surfden} can be rewritten as
\begin{equation}\label{eq:alt1}
    M=\Sigma\cdot\ area= \Sigma\cdot 2RL = \frac{\Sigma}{A}\cdot L^2,
\end{equation}
where in order to fulfil Eq.~\ref{eq:ML} it requires that
\begin{equation}\label{eq:alt2}
    \frac{1}{a^2}=\frac{\Sigma}{A}.
\end{equation}
On the other hand, the average density can be recalculated as
\begin{equation}\label{eq:alt3}
    n=\frac{mass}{volume}=\frac{\Sigma\cdot 2RL}{\pi R^2 L}=\frac{4}{\pi}\cdot\Sigma\ A\ L^{-1},
\end{equation}
which again -- to satisfy Eq.~\ref{eq:nL} -- necessitates that
\begin{equation}\label{eq:alt4}
    \frac{4}{\pi}\cdot\Sigma\ A =b_0,
\end{equation}
which can be also reordered to get Eq.~\ref{eq:surfden_AR}.
Introducing Eq.~\ref{eq:alt4} into Eq.~\ref{eq:alt2} leads back to Eq.~\ref{eq:a_AR} or Eq.~\ref{eq:a_Sigma} depending whether $\Sigma$ or $A$ is replaced.

\section{Observational data}\label{sec:appendixB}

The large number of datapoints ($>20\,000$ filaments) hampers the visualization of the $\Sigma$ variations in Fig.~\ref{fig:prediction_N}.
In addition to this figure in the main text, in
Fig.~\ref{fig:prediction_N_3panels} we show the different groups of filaments displayed in four ranges of average column density ($\Sigma$) similar to those defined by our models (see legend). We note that each of these groups populates a different part of the M-L phase space depending on their corresponding $\Sigma$ following the predictions of our models (Sect.~\ref{sec:toymodel}).

\begin{figure}[h]
\centering
\includegraphics[width=1.\linewidth]{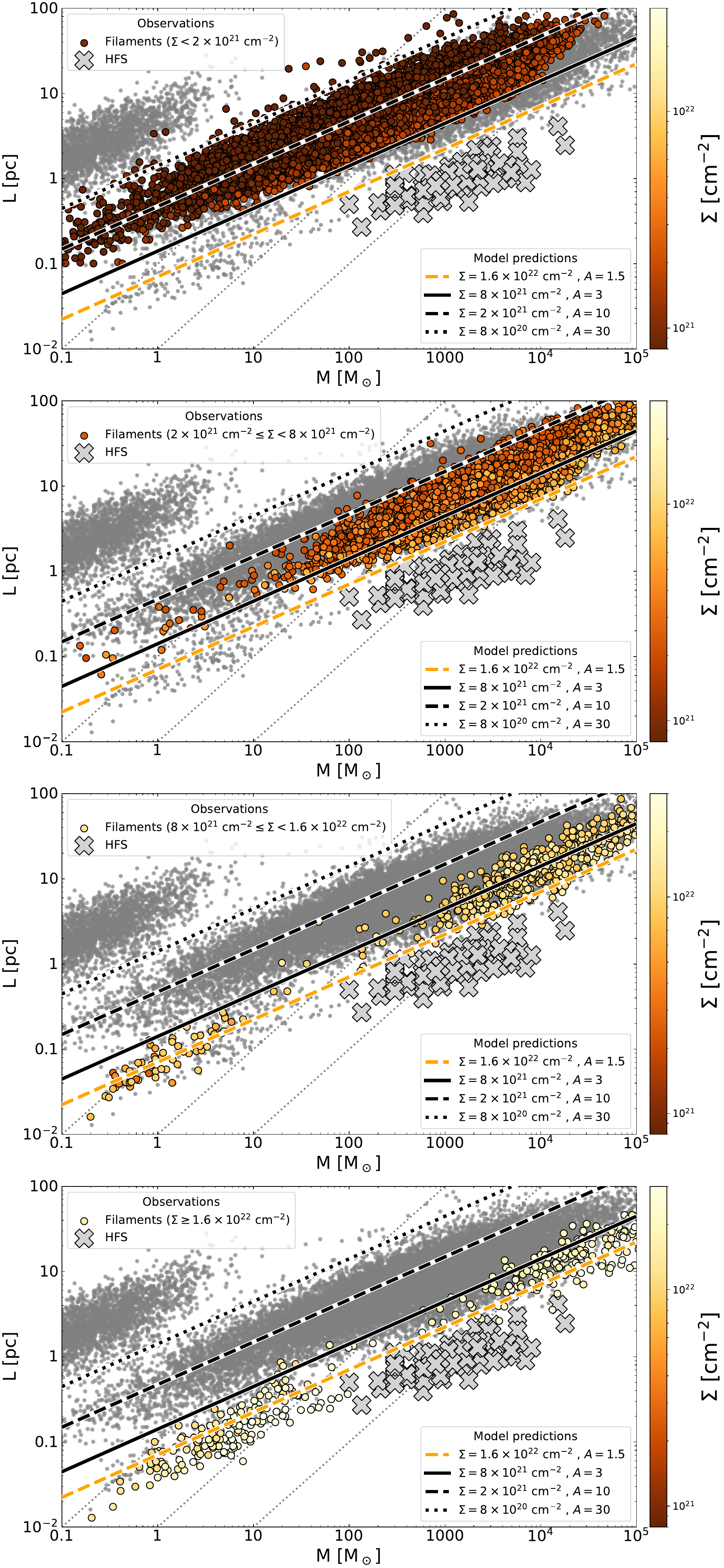}
      \caption{Same as Fig.~\ref{fig:prediction_N} but where we display different filament populations grouped in four ranges of average column density ($\Sigma$) similar to our models (see also legend). From top to bottom:
      {\bf (a)}  $\Sigma < 2\times 10^{21}$~cm$^{-2}$,
      {\bf (b)}  $2\times 10^{21}$~cm$^{-2} \leq \Sigma < 8 \times 10^{21}$~cm$^{-2}$,
      {\bf (c)}  $8\times 10^{21}$~cm$^{-2} \leq \Sigma < 1.6 \times 10^{22}$~cm$^{-2}$,
      and
      {\bf (d)}  $\Sigma \geq 1.6 \times 10^{22}$~cm$^{-2}$.
      }
\label{fig:prediction_N_3panels}
\end{figure}

\end{appendix}

\end{document}